\documentclass[12pt,preprint]{aastex}

\slugcomment{Revised \& Accepted for PASP: Dec 17, 2001}

\shorttitle{On the Nature of Stars with Planets}
\shortauthors{Reid}

\begin{document}
\title{On the Nature of Stars with Planets}

\author{I. Neill Reid\altaffilmark{1}}
\affil{Space Telescope Science Institute, 3700 San Martin Drive, Baltimore,
MD 21218; \\
Department of Physics and Astronomy, University of Pennsylvania, 209 South 33rd
Street, Philadelphia, PA 19104}
\email{inr@stsci.edu}

\begin{abstract}
We consider the metallicities and kinematics of nearby stars known to have planetary-mass
companions in the general context of the overall properties of the local Galactic Disk.
We have used Str\"omgren photometry to determine abundances for both the extrasolar-planet host
stars and for a volume-limited sample of 486 F, G and K stars selected from the Hipparcos catalogue.
The latter data show that the Sun lies near the modal abundance of the disk, with over 45\%
of local stars having super-solar metallicities. 
Twenty of the latter stars (4.1\%) are known to have planetary-mass companions. Using that
ratio to scale data for the complete sample of planetary host stars, we find that the fraction of stars
with extrasolar planets rises sharply with increasing abundance, confirming previous results. 
However, the frequency remains at the 3-4\% level for stars within $\pm0.15$ dex of
solar abundance, and falls to $\sim1\%$ only for stars with abundances less than half solar.
Given the present observational constraints, both in velocity precision and in the available time
baseline, these numbers represent a lower limit to the frequency of extrasolar planetary systems.
A comparison between the kinematics of the planetary host stars and a representative sample of disk stars
suggests that the former have an average age which is $\sim60\%$ of the latter.
\end{abstract}


\section {Introduction}

The discovery of the first extrasolar planetary system stands as one of
the key scientific and philosophical advances of the twentieth century. While
the existence of other planetary systems had been postulated for several 
centuries (Dick, 1998), and could even be regarded as likely, particularly 
following the detection of circumstellar disks around young stars (see 
Sargent \& Beckwith, 1993), Mayor \& Queloz' (1995) radial velocity
measurements of 51 Pegasi marked a definitive transition from speculation to
observation. The relatively short time interval which has elapsed since that
initial discovery has seen the identification of a plethora of additional
systems, notably by Marcy, Butler and collaborators.
Taken together, those systems provide sufficient numbers for a statistical
comparison of the characteristics of stars with planetary-mass companions 
against the overall distribution of properties of the local Galactic Disk. 
The results of such a study have obvious implications for estimating the
likely frequency of extrasolar planets (ESPs), particularly potentially habitable
systems.

Comparative studies of this type must pay due regard to several important
caveats. First, it is clear that most of the ESP systems discovered to
date bear little resemblance to our own Solar System: 51 Pegasi-like systems
feature `hot jupiters', Jovian-mass planets in sub-Mercurian orbits, while over
half of the current ESP catalogue have orbital eccentricities comparable to, or 
exceeding, that of Mercury and Pluto. Those circumstances, however, may at least partly reflect
observational selection; these systems have relatively short periods and relatively
high velocity amplitudes, and are therefore the easiest to detect. All of the `hot
jupiter' ESPs have reflex motions of tens of ms$^{-1}$, and it seems likely that we
have a fairly complete census of these objects. However,  it is only now
that observations are achieving both the requisite velocity precision and the 
decade-plus time baselines which are required for the detection of Jovian analogues,
and systems bearing a closer resemblance to the Solar System are starting
to emerge amongst the most recent discoveries (Vogt {\sl et al.}, 2001). Thus,
it is possible that the properties of the current ESP catalogue may reflect 
extreme, rather than characteristics, systems.

By the same token, it seems likely that the present catalogue includes only
a subset of extrasolar planetary systems in the Solar Neighbourhood. Studies estimate
that between 3 and 5\% of F, G-type stars have currently-detectable ESP systems
(Marcy \& Butler, 2000). Tabachnik \& Tremaine (2001), in particular, have used
maximum-likelihood analysis to estimate that current observations indicate a planetary
frequency of 3\% amongst solar-type stars, but that the frequency might be as high as 15\%
if the companion mass function is extrapolated to terrestial-mass systems.
Thus, the observed detection frequency may well underestimate the true frequency of solar-type 
stars with planetary systems, and possibly provides a biased
sampling of their characteristics.

Nonetheless, the current dataset offers a first cut at determining the conditions
required for the formation of planetary systems.  How are the ESP primaries 
distinguished from the average local field star? Studies to date have focused on
chemical abundance, with strong indications that stars known to have
planets tend to have solar or super-solar metallicity (Gonzalez, 1998; Santos et al, 2001).
While this may indicate a requirement on the initial conditions at formation, there
have also been suggestions that these higher abundances are a consequence of
planet formation (Lin {\sl et al.}, 1996), reflecting pollution of the stellar atmosphere by migrating
gas giants (Gonzalez, 1997; Laughlin, 2000).

Placing this result in the broadest context requires consideration
of both correlations which might exist with other properties of the
planetary host stars, and comparison against data for a reliable reference
sample of representative disk stars. The latter criterion is not met
in some recent analyses. In this paper we re-examine the abundance 
distribution of the ESP hosts, matched against similar data for an Hipparcos-based,
volume-limited sample of FGK stars. We also compare the kinematics
of ESP hosts against the velocity distribution of local disk stars. 
The paper is organised
as follows: the following section presents basic data for the ESP host stars;
section 3 discusses abundance calibration and the metallicity distribution;
section 4 examines the kinematics of the sample; and section 5 summarises
our main conclusions.

\section {The planetary hosts in the HR diagram}

\subsection {The sample}

Table 1 lists basic photometric and parallax data for stars currently
known to possess at least one planetary-mass companion. We shall
refer to those stars as ESP host stars.
In compiling this list, we follow the Geneva convention 
({\sl http://obswww.unige.ch/~udry/planet/}) of setting an upper mass limit of
M$_2\sin{i} = 17$M$_J$, where M$_J$ is the mass of Jupiter. There
are only four systems where M$_2\sin{i}$ exceeds 10M$_J$.
The parameters listed for the planetary systems are taken from 
the Extrasolar Planets Encyclopedia 
maintained by J. Schneider at {\sl http://cfa-www.harvard.edu/planets/}.

Since we only measure M$_2\sin{i}$ for most of these systems, there is clearly potential
for the inclusion of higher-mass companions on low-inclination orbits, either low-mass stars 
or brown dwarfs. Indeed, there may well be an overlap between the upper mass
range of planets and the lower mass-range of brown dwarfs\footnote{ In making this
statement, we define planets as forming in a circumstellar disk, while brown
dwarfs form as independent accreting cores in the parent molecular cloud. Following
common usage, we also require planets to be in orbit around a more massive central
body. Other definitions of these terms are possible.}, leading to an inherent 
ambiguity in interpretation. Since those two classes of objects may have different
intrinsic properties, it is important to conside the likely level of cross-contamination.

 The degree of contamination depends on the prevalence of brown dwarfs as close
companions to solar-type stars. Few unequivocal examples of such systems have been
detected, leading both to the postulation of a `brown dwarf desert' at $a < 10$ a.u.
(Marcy \& Butler, 2000), and support for the hypothesis that the low mass 
(M$_2\sin{i}<10$M$_J$) companions that {\sl are} detected are not a simple extension of
the companion mass function at higher masses. On the other hand, there have been
counter suggestions. Both Heacox (1999) and Stepinski \& Black (2000) have 
pointed out the similarity between the orbital properties of planetary-mass systems
and stellar binaries, although that might reflect similar dynamical evolution rather
than similar origins. More directly, Han {\sl et al.} (2000) and Gatewood {\sl et al.} (2000)
have analysed radial velocity data in tandem with Hipparcos Intermediate Astrometric Data,
and claim that in many cases the best-fit orbits have low inclination, and correspondingly
high true masses in the brown dwarf or even stellar r\'egime. 

The Han {\sl et al.} result has been scrutinised intently  and
generally found wanting. Statistically, the scarcity of systems with M$_2\sin{i} > 15$ M$_J$
demands a rather unlikely observational conspiracy, with orbital axes aligned within a few degrees
of the line of sight. Under a random distribution of inclinations, one would expect
several hundred systems with brown-dwarf mass companions for each planetary-mass system; not
only are the latter systems not observed, but there are not sufficient G dwarfs in the
Solar Neighbourhood to meet the numerical requirements. 
The amplitudes of the derived astrometric orbits are comparable with the uncertainties
in the Hipparcos IAD measurements, and  Pourbaix (2001) has shown that the low inclinations 
found by Han {\sl et al.} are largely an artefact of the fitting technique used. Pourbaix \&
Arenou (2001) conclude that $\rho$ CrB is the only system where the data merit interpretation
as a near face-on orbit, but HST astrometry  by McGrath {\sl et al.} (2001) sets an upper
limit on the semi-major axis at $a\sin{i} < 0.3$ milliarcseconds (mas) corresponding to M$_2 < 30 $M$_J$, 
rather than the 1.5 mas and $0.14\pm0.05$M$_\odot$ suggested by Gatewood {\sl et al.} (2000). 

Current estimates of the companion mass function are generally in good agreement, finding
approximately equal numbers per decade in mass (${dN \over dM} \propto M^{-1}$) for
$M < 10$M$_J$, with a sharp drop at higher masses (Jorissen {\sl et al.}, 2001; 
Zucker \& Mazeh, 2001; Tabachnik \& Tremaine, 2001). 
Indeed, astrometric observations have shown that
many of the candidate brown dwarf companions are, in fact, low-mass M dwarfs 
(Halbwachs {\sl et al.}, 2000), further enhancing the brown dwarf desert.
Based on these results, we expect little contamination ($<$5\%) in the ESP sample
listed in Table 1. Nonetheless, one star deserves comment.  As discussed later, HD 114762 
is one of the most metal-poor stars in the sample. The measured stellar rotational velocity suggests
that the star is being viewed close to pole-on, implying a low inclination for an 
equatorial orbit. 
In that case, the companion, detected by Latham {\sl et al.} (1989) is a candidate for the
first brown dwarf identification.

\subsection {The (M$_V$, (B-V)) colour-magnitude diagram}

All of the stars except
BD -10 3666 were observed by Hipparcos, and most have trigonometric parallaxes
measured to an accuracy of better than 5\%. The BV photometry listed in Table 1
is also taken from the Hipparcos catalogue (ESA, 1997), although we
use the literature data cited therein in preference to Tycho photometry,
given the systematic offset (and occasional large random errors) between the
latter system and standard Johnson data (Bessell, 2000).  BD -10 3666 has UBV
photometry by Ryan (1992) but no measured trigonometric parallax, and the
absolute magnitude listed in Table 1 is based on an estimated photometric parallax. 
Since both this star and HD 4203 were selected for observation based on the known
high metallicity, neither plays a role in the statistical comparison discussed in
the following sections.

The photometry and astrometry listed in Table 1 allow a detailed assessment of the
distribution of the ESP host stars in the HR diagram. Figure 1 makes that
comparison, where the reference main sequence is provided by data for stars
within 25 parsecs which have accurate photometry and trigonometric parallaxes
measured to a precision of ${\sigma_\pi \over \pi} < 10\%$. All of the latter
stars are members of the Galactic Disk. Three features are immediately noticeable:
\begin{enumerate}
\item G dwarfs contribute the majority of planet detections, with Gl 876 still
the only M dwarf known to have planetary-mass companions. To a large extent, the
distribution likely reflects the continuing observational focus on solar-type
stars.
\item The current sample includes a significant number of evolved stars. At least
six stars lie at the base of the subgiant branch, while four have evolved a considerable
distance from the main sequence. These stars are identified in Table 1.
\item The ESP host stars are distributed throughout the full width of the main
sequence. This is important since chemical abundance is the dominant factor
which governs the location of single stars on the main sequence - at a given
colour, metal-rich stars are more luminous than metal-poor stars. Thus, the observed
distribution points to a range of metallicity amongst stars with planets which is
at least comparable to the abundance distribution amongst the underlying local
disk (thin+thick) population.
\end{enumerate}

The last point is particularly pertinent given the recent emphasis laid on the
high metal abundance measured for at least some of the ESP primaries (Gonzalez {\sl et al.},
1999, 2001; Santos {\sl et al.}, 2001). Many of those abundances are
significantly higher than the value usually taken as the median for the Galactic
Disk. Most previous studies, however, treat the metallicity distributions
of the ESP hosts and of the disk as separate entities.
The following section places the former in the context of the latter, and
considers how the high chemical abundances are reconciled with
the distribution evident in Figure 1.

\section {Chemical abundances}

\subsection{Measuring stellar metallicity}

The metal content of stellar atmospheres can be measured using a wide variety of
techniques. In general, the accuracy of the final measurement is at least
inversely proportional to the difficulty of the observation. Analyses based on
high-resolution spectroscopy are usually more reliable than those which 
utilise broadband colours, but photometric data are obtained much more readily 
than echelle spectra. Thus, any statistical analysis requiring a dataset of
even modest dimensions must balance two factors - availability and accuracy.

As a further complication, comparative studies must
ensure that data drawn from different analyses are tied to a consistent system.
Different measurement techniques not only have different random uncertainties,
but can also exhibit systematic discrepanciesin scale and/or zeropoint, 
as discussed in the context of 
globular cluster distance determination by Gratton {\sl et al.} 1997) and Reid (1998).
Apart from differences in the choice of standard stars, we note that the solar
iron abundance was re-calibrated relatively recently, revised downward from A(Fe)=7.67 to
A(Fe)=7.54 (Biemont {\sl et al.}, 1991), where A(Fe) is the logarithmic abundance
on a scale where A(H)=12. While most abundances are measured differentially, this 
re-calibration might lead to a systematic offset depending on how and when the
abundances of the standard stars were determined. 

This issue is a concern since, while nearly all of the ESP host stars have recent
high-resolution spectroscopic abundance measurements, most estimates of the
underlying field-star metallicity distribution rest on lower resolution techniques. 
Santos {\sl et al.} (2001) have addressed this problem to some extent by providing
high-resolution spectral analyses for 43 G dwarfs in 42 systems drawn from the
volume-limited sample of stars with (B-V)$< 1.1$ and $d < 17$ parsecs. That
sample is scarcely sufficient in size, however, to provide an adequate mapping of
the distribution of disk properties. We adopt an alternative strategy in this paper.

Str\"omgren uvby photometry provides a relatively simple means of determining
abundances for F, G and early K-type stars (Str\"omgren, 1966). 
Data are available for most solar-type stars brighter than 10th magnitude.
Indeed, both Gim\'enez (2000) and Laughlin (2000) have applied those
measurements to studying the ESP host stars.
Metallicities are determined by measuring the differential line-blanketting via
the m$_1$ and c$_1$ indices, where
\begin{displaymath}
m_1 \ = \ (v-b) \ - \ (b-y) 
\end{displaymath}
and
\begin{displaymath}
c_1 \ = \ (u-v) \ - \ (v-b)
\end{displaymath}
The latter index is also gravity sensitive, allowing discrimination between
dwarfs and subgiants. 

The literature contains several calibrations of the Str\"omgren indices
against metallicity. Gim\'enez adopts that given by Olsen (1984); we follow
Laughlin in using the more recent calibrations derived by Schuster 
\& Nissen (1989). They provide two calibrating relations,
\begin{eqnarray*}
[Fe/H]_{uvby} & = & 1.052 - 73.21m_1 + 280.9m_1(b-y) + 333.95m_1^2(b-y) - 595.5m_1(b-y)^2 \\
 & & + [5.486 - 41.61m_1 - 7.963(b-y)] \times \log{(m_1 - [0.6322 - 3.58(b-y) + 5.20 (b-y)^2])}
\end{eqnarray*}
for F stars, ($0.22 \le (b-y) < 0.375$, $0.03 \le m_1 \le 0.21$, $0.17 \le c_1 \le 0.58$
and $-3.5 \le [Fe/H] \le 0.2$), and
\begin{eqnarray*}
[Fe/H]_{uvby} & = & -2.0695  - 22.45m_1 - 53.8m_1^2 - 62.04m_1(b-y) + 145.5m_1^2(b-y) \\
 & & (85.1m_1 - 13.8c_1 - 137.2m_1^2)c_1 
\end{eqnarray*}
for G stars ($0.375 \le (b-y) \le 0.59$, $0.03 \le m_1 \le 0.57$, $0.10 \le c_1 \le 0.47$
and $-2.6 \le [Fe/H] \le 0.4$). 

Apart from BD -10 1366 and Gl 876, where Str\"omgren data offer little useful
information, only HD 177830 (HIP 93746) lacks uvby photometry. Table 2 lists (b-y), m$_1$ and
c$_1$ colour indices, taken from Hauck \& Mermilliod's (1998) catalogue, and the
resulting [Fe/H]$_{uvby}$ for the remaining stars, together with the metallicities 
derived from high-resolution spectroscopy. As expected, there is 
a systematic offset to lower abundances in the Str\"omgren calibration. 
The results are compared in Figure 2, where the upper panels plot the full dataset
(see also Figure 1 in Laughlin, 2000).  We note that residuals
tend to increase among early-type K stars, (b-y)$>$0.5, where both methods become
more problematic. 

We have quantified the comparison using the datasets from Santos {\sl et al.} (2001) and
Gonzalez and collaborators, which provide two internally consistent datasets of 
moderate size. Based on the twenty-two ESP host stars observed by Santos {\sl et al.} (2001), we
derive
\begin{displaymath}
\delta [Fe/H] \ = \ [Fe/H]_{uvby} - [Fe/H]_{sp} \ = \ -0.117\pm0.095
\end{displaymath}
where the uncertainties quoted are  the rms dispersion about the mean. The formal
standard error of the mean, $\sigma_\mu$,  is 0.020 dex and the median offset is -0.13 dex.
The twenty-two stars in the Gonzalez dataset give an almost identical result,
\begin{displaymath}
\delta [Fe/H] \  = \ -0.118\pm0.113, \quad \sigma_\mu \ = \ 0.024 
\end{displaymath}  
The median offset is -0.10 dex.
Combining both datasets with forty field stars from the Santos {\sl et al.} reference
sample gives
\begin{displaymath}
\delta [Fe/H] \  = \ -0.083\pm0.118,  \quad \sigma_\mu \ = \ 0.013 
\end{displaymath} 
The median offset is -0.07 dex.
This is in excellent agreement with Gratton {\sl et al.'s} independent analysis of
152 stars spanning a much larger range of metallicity, where they derive
\begin{displaymath}
\delta [Fe/H] \  = \ -0.102\pm0.151, \quad \sigma_\mu \ = \ 0.012
\end{displaymath}
Given uncertainties of $\pm0.06$ dex in the Santos {\sl et al.} and Gonzalez 
spectroscopic measurements, the measured
dispersion indicates that the Str\"omgren data have typical
uncertainties of $\pm0.1$ dex at near-solar abundances, sufficient accuracy for present purposes.

Based on this comparison, we conclude that the Str\"omgren abundance scale
is offset by -0.1 dex from the most recent calibrations.  
We note that the offset is intriguingly similar to the re-calibration of the
solar abundance, although that similarity may be coincidental.
Rather than attempt to correct the metallicity measurements, we base our
analysis on the {\sl uvby} scale; in effect, we adopt the somewhat paradoxical
definition 
\begin{displaymath}
[Fe/H]_{uvby} (\odot) \ = \ -0.1
\end{displaymath}
Since we are considering the comparative distributions of field star and 
ESP host star abundances, consistency is more important than the
numerical value chosen for the fiducial zeropoint.

\subsection {The reference sample}

Defining a suitable reference sample is crucial to assessing how the 
properties of the ESP host stars map onto the overall field star distribution.
Complete, volume-limited datasets offer the most reliable comparison, but 
have generally not been available to previous studies. Murray {\sl et al.} (2001)
have attempted to turn Hipparcos data to this end, with a reference sample defined 
by selecting HD stars with $\pi > 10$ mas and ${\sigma_\pi \over \pi} < 10\%$.
However, the colour-magnitude diagram for this dataset (Figure 1 in Murray \& Chaboyer, 2001)
is clearly biased strongly towards F-type and early G-type dwarfs, partly
reflecting sampling in the Hipparcos catalogue at $V > 9$th magnitude, and 
partly reflecting the fact that the
HD catalogue was selected from blue photographic plates. The resulting dataset therefore
provides a biased subset of nearby disk stars. Murray {\sl et al.'s} metallicities are taken
from the Cayrel de Strobel (1997) catalogue, and therefore represent an amalgam of heterogenous
sources. Thus, this dataset is not suitable as
a local reference. 

Several other analyses 
(e.g. Gonzalez, 1999; Butler {\sl et al.}, 2000) have relied on the
nearby-star sample defined by Favata {\sl et al.} (1997: F97) to represent
the abundance distribution of local disk stars. 
That sample, however, is severely flawed in several important respects, as 
illustrated in Figure 3. The F97 abundances are derived
from high-resolution spectra, and a comparison with Str\"omgren abundances
gives
\begin{displaymath}
\delta [Fe/H] \ = \ -0.05\pm0.15, \ {\rm 69 \ stars}
\end{displaymath}
somewhat less than the offsets derived in the previous section. The full sample
of 90 stars includes both components of several binaries, giving those
systems double weight in the abundance distribution, and spans a substantially
larger colour range than ESP host stars (excluding Gl 876). Finally, and 
most significantly, the sample is neither complete nor volume-limited. Favata {\sl et al.}
(1996) constructed the original sample by taking a randomly-selected subset of 200 stars 
with $0.5 < (B-V) < 1.4$ from the second Catalogue of Nearby Stars (Gliese, 1969; 
Gliese \& Jahrei{\ss}, 1979: CNS2). Ninety-four of those stars were observed spectroscopically.
Unfortunately, while the CNS2 has a nominal distance limit of 25 parsecs, subsequent
Hipparcos astrometry has shown that a significant fraction of the stars lie at much
larger distances. Figure 3 shows that at least 40\% of the F97 dataset lies beyond
25 parsecs.
Our analysis demands a more reliable reference dataset.

Figure 4 shows the (M$_V$, (B-V)) colour-magnitude outlined by the 1549 stars in the
Hipparcos catalogue with formal trigonometric parallax measurements exceeding 40
mas ($d \le 25$ parsecs, (m-M)$\le 1.99$). As in Figure 1, we use the
literature BV photometry in the catalogue in preference to Tycho data. We
have not applied any cut based on parallax precision, but use different symbols to identify 
the 1477 stars with parallaxes measured to a precision of better than 20\%. 
Nearly all of the
stars lying below the main-sequence in this figure have inaccurate parallax
measurements. 

The box superimposed on Figure 4 isolates 488
stars with $0.5 \le {\rm (B-V)} \le 1.0$ and $2.0 \ge {\rm M}_V \ge 7.0$, 
matching the colour/magnitude range of the bulk of the ESP host stars. 
The limiting magnitude for completeness in the Hipparcos catalogue is 
\begin{displaymath}
V \ = \ 7.9 \ + \ 1.1 sin|b|
\end{displaymath}
so the 25-parsec sample is effectively complete over the whole sky 
for M$_V \le 5.9$ (the dotted line in Figure 4). The 
full catalogue is $\sim25\%$ incomplete for stars with $8 \le {\rm V} < 9$, but
should be significantly more complete for nearby stars, since all stars
suspected of being within 25 parsecs were included in the
input catalogue - as emphasised by the large numbers of M dwarfs in Figure 4.
Indeed, Jahrei{\ss} \& Wielen (1997) argue that the Hipparcos catalogue is essentially
complete to M$_V=8.5$ for stars within 25 parsecs of the Sun.  
Thus, the F, G and K stars isolated in Figure 4 effectively represent a
complete, volume-limited sample. We identify these stars as the FGK25
Hipparcos dataset. Note that the sample includes 20 known ESP hosts from Table 1.

As noted in the previous section, Str\"omgren photometry is now available for 
a substantial number of bright F, G and K stars, and is readily accessible through
the catalogue compiled by Hauck \& Mermilliod (1998). We have cross-referenced
the FGK25 Hipparcos dataset against that catalogue and located photometry for 419
of the 486 stars - 86\% of the sample. 
Those stars are identified as solid squares in Figure 4. The sample includes
members of binary systems, but only one star per system.
It is clear that almost every star with (B-V)$<$0.8 has
photometry, while the late-G and K dwarfs which lack Str\"omgren measurements
are distributed over the full width of the main sequence, interspersed with 
stars which have uvby photometry. We conclude that 
the abundance distribution deduced 
from these data is characteristic of the parent population(s) of the ESP
host stars. 

\subsection {The abundance distribution of local field stars}

The overwhelming majority of the stars in the FGK25 sample are members of
the Galactic Disk.
Five stars, however, lie significantly below the main sequence. 
These are HIP 57939, 62951, 67655, 79537 and 79979. Two of these stars,
HIP 62951 and HIP 79979, are in binary systems where the companion has
affected the parallax determination; Fabricius \& Makarov (2000) have
reanalysed the Hipparcos data for HIP 62951 and find $\pi=2.4$ mas. 
These are the only two stars in the FGK25 sample with ${\sigma_\pi \over \pi} > 0.2$
(there are only 6 other stars with ${\sigma_\pi \over \pi} > 0.1$), 
and we exclude both from our sample.
The remaining three subluminous stars  are {\sl bona-fide} metal-poor subdwarfs.
HIP 57939 is HD 103095, or Groombridge 1830, the well-known intermediate-abundance 
([Fe/H]$_{uvby}$=-1.4) subdwarf; 
HIP 67655, or HD 120559, has [Fe/H]$_{uvby}$=-0.94; and HIP 79537 is HD 145417,
[Fe/H]$_{uvby}$=-1.25. The presence of three such stars in this volume-limited sample
suggests a somewhat higher density normalisation ($(\sim0.6\pm0.35)\%$ relative to the disk
population) than
usually adopted for the local Galactic halo\footnote{This has implications for the
interpretation of the nature of the cool white dwarfs identified in 
recent proper motion 
surveys (Oppenheimer {\sl et al.}, 2001). An increased local density 
of the {\sl stellar} halo easily accounts for the observed numbers of
high velocity stars, as suggested by  Reid {\sl et al.}, 2001).}.

Before comparing these results against other recent analyses, we should emphasise the limited
nature of the present study. The question addressed here can be stated as follows: \\
Based on current statistics, and given a sample of stars drawn from the Galactic mid-Plane near 
the Sun, what is the frequency of ESP systems as a function of metallicity. 

Our goal in constructing the reference sample of field stars, therefore, is not an unbiased
estimate of the present-day metallicity distribution of the Galactic Disk - a parameter
used to constrain Disk star formation histories. 
That undertaking requires
limiting analysis to stars with main-sequence lifetimes older than the age of the Disk, 
avoiding possible bias through a disproportionate contribution
from  recent star formation episodes.
Moreover, given a potential correlation between metallicity and velocity, one should weight each
star's contribution by its W-velocity to allow for the residence time in the mid-Plane and 
convert volume density to surface density. The latter is not an option for the present sample,
since over 25\% of the stars lack radial velocities. Moreover, our (B-V) limits, modelled
on the known ESP systems,  include
early-type G and late-F stars, whose main-sequence lifetimes are shorter than 10 Gyrs.
Thus, our field star sample is tailored to provide
a local snapshot of the present-day abundance distribution in the mid-Plane, rather than
an integrated history of star formation in the Disk. 

We have compared the abundance distribution of the FGK25 dataset against results from 
two other studies: the Favata {\sl et al.} (1997) analysis, described above, and the
recent study by Haywood (2001: H2001). In both cases, we consider volume-limited samples
(i.e. the distribution is not weighted by W velocity as in Figure 3 of F97). 
Haywood's analysis {\sl is} aimed at determining an unbiased estimate of
the Disk abundance distribution, so while his initial sample is selected  to have
M$V < 8.5$, (B-V)$>$0.25 and $\pi > 40$ mas (based on Hipparcos data), the 
final analysis is limited to 328 stars with
$M_V > 4.5$; that is, main-sequence dwarfs with lifetimes longer than the age of the Disk. 
Abundances are derived primarily from Geneva photometry, supplemented by Str\"omgren data, 
with the metallicity scale effectively adjusted to the high-resolution (Santos/Gonzalez) system. 

Figure 5 compares the abundance distributions derived in those studies against our own results. 
For consistency, we have adjusted all of the metallicity scales to match [Fe/H]$_{uvby}$.
All three distributions peak at values close to the solar abundance, with a substantial fraction
of the sample (45\% in the FGK25 dataset) having super-solar metallicities. As discussed
by Haywood, this represents a significant revision of previous analyses, and may reflect
a bias against metal-rich stars in samples selected based on spectral type rather than
distance/colour. We note that only a small fraction of the local Disk have abundances of less than
$1 \over 3$ solar: only 25 stars (6\%) in the FGK25 sample.  If
$\approx10\%$ of the local stars are members of the thick disk,
as suggested by kinematic analyses (Reid, Hawley \& Gizis, 1995), then
the mean abundance of that sub-population lies much closer to the solar metallicity than
the value of [Fe/H]=-0.6 adopted in some Galactic models.

Both the F97 and H2001 samples show a more extended distribution towards higher abundances than
the FGK25 dataset. This is somewhat surprising, since the latter sample, extending to
 stars brighter than M$_V=4$, 
should include a higher proportion of younger stars which are likely to be more metal-rich.
The discrepancy may originate from the abundance calibrations. Figure 6 shows the metallicity
distribution as a function of (B-V) colour for the H2001 and FGK25 datasets. The former
shows a clear trend of increasing metallicity at redder colours, suggesting a possible
systematic bias in the abundance calibration of Geneva photometry. Further observations
are required to verify this hypothesis.

For present purposes, the most significant point is that the metallicities discussed  here for both
field stars and ESP host stars are derived from a single source - Str\"omgren photometry,
as calibrated by Schuster \& Nissen (1989). Thus, the comparison between the
two abundance distributions described in the following section is internally fully
self-consistent.

\subsection { The abundance distribution and the frequency of giant planets}

Before comparing metallicities, Figure 7 matches the (m$_1$, (b-y)) and (c$_1$, (b-y))
distributions of the ESP host stars and the FGK25 Hipparcos sample. We have distinguished between
main sequence stars and potential subgiants in the former sample. As previously
noted by Gim\'enez (2000), a significant fraction of the ESP host stars have high c$_1$ values, 
suggesting low gravities and a mildly evolved status. In some cases, however, the colours reflect
high metallicity rather than low gravity; thus, both HIP 43587 (55 Cnc) and 79248 (HD 145675) lie
well above the ((b-y), c$_1$) sequence at (b-y)$\sim$0.42, but their location in the
(M$_V$, (B-V)) plane demands that both are main-sequence dwarfs. Spectroscopy indicates that
both are super-metal rich (Table 2).  Nonetheless, the fraction of 
subgiants amongst known ESP hosts ($\sim15\%$) is at least a factor of three higher than
that in the volume-limited sample. This may reflect an observational selection effect, since
evolved stars are intrinsically more luminous, and therefore more likely to be
included in the radial velocity monitoring programs. 

 The reddest star
in the sample is the K2 subgiant, HD 27442 (HIP 19921). This lies beyond 
the formal limits of the Schuster \& Nissen (1989) abundance calibration, at (b-y)=0.65; however, the
derived metallicity, [Fe/H]$_{uvby}$=0.26, is not inconsistent with the spectroscopic
determination of [Fe/H]=+0.22, so we have retained the star in the sample. 
We have excluded both BD -10 1366 and HD 4203 from the statistical comparison.
As noted above, those stars were added to the Keck/Lick radial velocity program because
they were known to be extremely metal-rich. The M dwarfs, Gl 876, is also excluded, 
but we include HD 177830, adjusting its abundance from [Fe/H]=0.36 to [Fe/H]$_{uvby}$=0.26.

The sample of ESP host stars is not volume-limited, particularly
given the fact that the original target list was constructed before the
availability of Hipparcos data, and is therefore subject to the type of distance
errors illustrated in Figure 3. There may therefore be underlying biases reflecting 
the initial selection of which stars to monitor. Those effects can be quantified once the
full dataset is available. 
Nonetheless, it is not unreasonable to hope that the current catalogue of ESP host stars 
provides a representative subset of stars with currently-detectable planetary systems;
that is, stars with relatively massive (super-Jovian) companions on relatively
short-period ($<$ few years) orbits.

Figure 8 plots the abundance distribution of both the ESP host stars and the 
reference FGK25 Hipparcos dataset. The subgiant contribution to the former
distribution is shown as the shaded histogram and is consistent with the
overall distribution. As emphasised in \S3.1, these data are all on the 
Str\"omgren system, [Fe/H]$_{uvby}$, placing the Sun as -0.1 dex, at the
mode of the local abundance distribution. 

Visual comparison clearly confirms previous suggestions that  
the abundance distribution of the ESP host stars is weighted more
heavily towards super-solar metallicity than the field distribution. 
To quantify that comparison we have combined the distributions
by scaling the upper distribution to match the observed fraction of
ESP hosts amongst the volume-complete sample. As noted above, 20 of the 486
stars in the FGK25 sample, or 4.1\%, are known to have planetary-mass
companions. This fraction is broadly consistent with previous estimates
(e.g.  Marcy \& Butler, 2000). We have used this factor to scale the 
metallicity distribution in the uppermost panel of Figure 8, and the 
two lowest panels in Figure 8 show the fraction of ESP host stars,
\begin{displaymath}
f_{ESP} \ = \ {N_{ESP} \over N_{tot}}
\end{displaymath}
as a function of metallicity. 
There is an obvious trend with
abundance, with f$_{ESP}$ rising to near unity at the highest abundances;
both of the stars in the highest-metallicity bin of the FGK25 sample are known
to have planetary-mass companions. However, 
even at an abundance of $2\over5$th solar, $\sim1\%$ of F, G and early-K stars
are predicted to have Jovian-mass planetary companions. HD 114762b, the likely
brown dwarf, contributes the spike at [Fe/H]$_{uvby}$=-0.7 dex.

Finally, we have compared the properties of the individual planetary systems against
the abundances derived from the Str\"omgren data. Figure 9 shows the results, where we
identify separately systems with multiple components (including the Sun, represented by Jupiter
and Saturn). There is no obvious correlation between [Fe/H]$_{uvby}$ and any of the
observed characteristics. 

\subsection {Discussion}

Is the correlation with metallicity evident in Figure 8 a selection effect?
Metal-rich stars are more luminous than their metal-poor counterparts, and
therefore, like subgiants, might be expected to be better represented in a
target list which is partly magnitude limited. However, it seems unlikely that
this type of bias could account for the smooth trend evident in the observations, 
particularly given the identification of planetary companions to BD -10 1366
and HD 4203, stars specifically added to the Keck program because they were
known to be super-metal-rich. Thus, the simplest interpretation of 
Figure 8 is that the correlation represents a real physical phenomenon.
The explanation for this phenomenon is somewhat less clear.

Perhaps appropriately, the two mechanisms proposed to account for the
observed correlation mirror the classic nature {\sl versus} nurture debates of
biological behavioural sciences. Under the first hypothesis, planetary systems
(at least those with giant planets) form more readily in the dustier environment 
likely to be present in high-metallicity circumstellar disks. Under the
alternative hypothesis, gas giants migrate inwards due to dynamical friction with
residual disk material and are absorbed into the stellar envelope.
The enhanced metal content of the planet (solar system giants are likely to have
Z$> 0.1$) enriches the metallicity of the outer convective envelope, leading to a higher 
measured chemical abundance. 

One question mark hanging over the planetary pollution hypothesis centres on the
details of the enhanced metal content of Jovian planets. In astronomical terms, 
`metals' encompass all elements except hydrogen and helium - but not all metals
are created equal. Metallicity measurements for F, G and early K-type stars are
primarily measuring blanketting due to heavy elements, notably iron. The giant
planets are known to have non-cosmic inter-element abundance ratios, but if the
additional `metals' are ices (C, N, O) rather than minerals (Fe, Si, Ni), as suggested 
by the possible absence of a rocky core in Jupiter (Guillot, 1999), then planetary
pollution will have little effect on the apparent metallicity of the stellar envelope. 

These two competing scenarios are discussed extensively by, amongst others,  Gonzalez 
(1997, 1999), Laughlin (2000), Murray {\sl et al.} (2001) 
and Santos {\sl et al.} (2001). A major prediction of 
the pollution hypothesis is that the degree of metallicity enhancement should increase
with increasing mass of the parent star. This follows from the corresponding decrease in 
mass of the convective envelope; adding high-Z material gives a proportionately larger 
increase in metallicity. Both Laughlin and Murray \& Chaboyer (2001) have argued that this effect 
is present, although the latter authors note that a similar trend is present in their 
reference sample, and both analyses are based on a subset of the current catalogue of ESP hosts.
Santos {\sl et al.} (2001), in contrast,  arrive at the opposite conclusion based on
analysis of more than 60 systems. In their analysis, they rightly place more emphasis on the 
location of the upper envelope of the abundance distribution as a function of mass, 
rather than the mean metallicity.

All three of these analyses 
use theoretical tracks to deduce masses for individual ESP host stars. Figure 10
shows an alternative, more empirical approach. We have separated the current sample into
main-sequence stars and subgiants based on location in Figure 1, 
and plot the abundance as a function of (b-y). 
For the main-sequence sample, (b-y) effectively traces mass, and the absence of
any strong trend in the location of the high-metallicity boundary, in particular,
a decrease in [Fe/H]$_{max}$ at redder (b-y) colours, supports
the conclusion reached by Santos {\sl et al.} Moreover, the latter authors point out that
as stars evolve onto the subgiant branch, the convective envelope increases in size, diluting 
the effect of any planetary pollution. It is clear from Figure 10 that the evolved 
stars can be as metal-rich as the main-sequence dwarfs; indeed, the
K2 subgiant HD 24427 (HIP 19921) is amongst the most metal-rich stars
in the sample.  Thus, these results suggest that planetary systems are born metal-rich,
rather than having high metallicity thrust upon them.

\section {Kinematics}

\subsection {Velocity dispersion and ages}

The age distribution of the ESP host stars is clearly an important parameter for
understanding the Galactic origins of these systems. Two techniques have been used to
estimate ages for individual stars: isochrone fitting; and the level of chromospheric
activity, as measured through emission at the Ca II H \& K lines. Both methods can
be applied to individuals, but both have limitations. Isochrone fitting provides reliable
ages for relatively F and early-G stars, but becomes less accurate for longer-lived,
later-type stars. Chromospheric ages, quantified using the $R'_{HK}$ index (Soderblom
{\sl et al.}, 1991), are more readily derived, but are also less reliable since there
is a considerable dispersion in activity amongst individual stars with similar
ages (see Figure 10 in Soderblom {\sl et al.}). Moreover, variability is an issue; 
as Henry {\sl et al.} (1996)
point out, the Sun's age could be estimated as anywhere between 2.2 and 8 Gyrs
depending on when the observations are taken during the Solar cycle. Finally, both
of these techniques become significantly less reliable at ages exceeding $\sim2$ Gyrs
(as evidenced by continuing uncertainties in the Galactic star formation history).

Nonetheless, both of these methods provide useful insight into the age
distribution, and both have been applied by Gonzalez and co-workers (Gonzalez \& Laws, 1998;
Gonzalez, 1999; Gonzalez {\sl et al.}, 2001) to estimate ages for 33 of the
systems listed in Table 1. A comparison between the different estimates emphasises the
inherent uncertainties, most dramatically for HD 217107 and HD 222582, where both have
chromospheric age estimates of 5.6 Gyrs, but isochrone estimates of 1.2 and 11 Gyrs, 
respectively. Averaging the results for all 33 stars, 
we derive a mean age of 5.6 Gyrs ($\sigma=3.6$ Gyrs).

Space motions cannot be used to provide age estimates for individual stars. However, 
stellar kinematics offer an alternative means of comparing the {\sl average} properties of
diverse groups of stars. Velocity dispersion increases with age, probably through
the mechanisms of orbital diffusion (Wielen, 1977) and scattering due to molecular
clouds (Spitzer \& Schwarzschild, 1953). A comparison between the velocity
distributions of the ESP host stars and the local disk can test whether there is a
significant difference in the mean age of the two samples. 

We have calculated space motions for the ESP hosts using astrometric data from
the Hipparcos catalogue and the available radial velocity measurements. Table 3 lists those
data and the resulting (U, V, W) motions, where U is positive toward the Galactic Centre, 
V positive in the direction of rotation, and W directed toward the NGP. 

All of the FGK25 stars have accurate proper motions and parallaes
from Hipparcos, but only 60\% have published radial velocities, rendering the sample unsuitable as a reference. 
However, the volume-complete M-dwarf sample from the PMSU survey of nearby stars (Reid {\sl et al.}, 1995: PMSU1)
gives a ready alternative, providing an unbiased representation of the kinematics of local disk
stars. Reid {\sl et al.} (2002: PMSU4) have revised the original dataset to incorporate more recent 
astrometric data, notably from Hipparcos, besides including higher-accuracy radial velocities
from echelle observations summarised by Gizis {\sl et al.} (2001: PMSU3). The final sample is
comparable in size to the FGK25 dataset, with 436 systems lying within M$_v$-dependent distance
limits ranging from 10 to 20 parsecs. 

Figure 11 compares the velocity distributions of the two datasets. The left-hand panels plot the
two-component velocity distributions; the right-hand panels show probability plots of the (U, V, W)
distributions. As originally discussed by Lutz \& Upgren (1980), these diagrams plot the
cumulative distribution of a sample, C(x), against the difference with respect to the mean value, $\bar x$,
in units of the standard deviation. 
A normal distribution, $f(x) \ = \ {1 \over \sqrt{2 \pi \sigma}} . e^{-{{(x-\bar x)}^2 \over \sigma^2}}$,
gives a straight line, slope $\sigma$, in this plane. Figure 11 plots three empirical velocity
distributions: data for the ESP host stars; for the full PMSU M-dwarf sample; and for the PMSU 
dMe dwarfs, with H$\alpha$ emission exceeding 1\AA\ equivalent width. As discussed by 
Hawley {\sl et al.} (1996; PMSU2), chromospheric emission is an age-dependent phenomenon, so the
last dataset is characteristic of a moderately young stellar population, $\langle \tau \rangle \approx 1-2$Gyrs.

It is clear from Figure 11 that the velocity distribution of the ESP host stars is more
closely  matched to the full M dwarf sample than to the dMe sample. Quantitatively, 
linear fits to the central regions of the probability plots ($-1.9 < rms < 1.9$)
give 
\begin{eqnarray*}
(\sigma_U  & = &  35.2;  \sigma_V  =  22.9;  \sigma_W  =  17.3:  U_\odot  =  -6.3; 
 V_\odot = -24.2;  W_\odot = -7.7) \\ 
& & \sigma_{tot}  =  45.4 \ {\rm km s^{-1}; 63 \ systems}
\end{eqnarray*}
where $\sigma_{tot}$ is the overall velocity dispersion.
Applying the same technique to the M dwarf samples gives
\begin{eqnarray*}
(\sigma_U & = & 21.2;  \sigma_V = 14.1;  \sigma_W = 13.0: U_\odot = -14.5; 
 V_\odot = -11.5; W_\odot = -8.1) \\ 
& &\sigma_{tot}  =  28.6 \ {\rm km s^{-1}; 69 \ systems}
\end{eqnarray*}
for the emission line dwarfs and
\begin{eqnarray*}
(\sigma_U & = & 39.1;  \sigma_V = 38.8;  \sigma_W = 23.6:  U_\odot = -5.7; 
 V_\odot = -9.6;  W_\odot = -3.3) \\
& & \sigma_{tot}  =  59.9 \ {\rm km s^{-1}; 404 \ systems}
\end{eqnarray*}
for the full sample. 

Based on this comparison, we conclude that the current sample of F, G and K-type ESP hosts
is younger, on average, than the overall disk population, but
includes stars significantly older than typical of the dMe sample. Quantitatively, if we assume
diffusion with $\sigma \propto \tau^{1 \over 2}$, then 
\begin{displaymath}
\langle \tau_{ESP} \rangle \ \sim \ 0.6 \langle \tau_{dM} \rangle \ \sim 2.5 \langle \tau_{dMe} \rangle
\end{displaymath}
suggesting an average age of 3-4 Gyrs for ESP host stars for an approximately uniform star-formation
rate in a 10-Gyrs-old disk. This younger mean age is not unexpected, given the
higher proportion of metal-rich stars amongst the ESP sample.
The average metal abundance of the Galactic disk is
expected to increase with time, as successive generations of star formation contribute additional
nucleosynthetic debris to the interstellar medium, so a sample biased toward high metallicities
is also likely to be biased towards stars that are younger than average. 

\subsection{Kinematics and metallicity}

The previous section considered the overall distribution of velocities of the ESP host stars.
We can also look for correlations using velocities for the individual stars, correcting the observed
heliocentric data for the solar motion with respect to the Local Standard of Rest (LSR). For the
latter parameter, we use
the values derived by Dehnen \& Binney (1998), 
\begin{displaymath}
( U_\odot, \ V_\odot, W_\odot; 10.0, \ 5.3, \ 7.2)
\end{displaymath}
where these values give the motion of the Sun with respect to the LSR. Thus, the Sun is moving
towards the Galactic Centre, towards the direction of rotation and towards the NGP, and the
observed velocities must be corrected accordingly. We denote the corrected velocities as
 (U', V', W').

Figure 12 plots velocities for the ESP host stars as a function of abundance. We
also indicate the location of the Sun on these diagrams. There is no obvious correlation between
metallicity and either the (U', V', W') component velocities or the total motion with
respect to the LSR, V$_{tot}$. The highest metallicity stars in the sample span essentially
the same range of velocities as the solar-abundance and sub-solar abundance ESP host stars.

Several previous studies have commented on the relatively low velocity ($\sim 13$ kms$^{-1}$) 
of the Sun with respect
to the LSR. Gonzalex (1999), in particular, has invoked the Weak Anthropic Principle (Barrow \& Tipler, 1988) 
in conjunction with this property, arguing that the small offset from co-rotation minimises
excursions into the potentially dangerous environment (supernovae, gravitational interactions) of 
spiral arms, therefore providing the long term quiescence which may be necessary for advanced
life forms to develop. We can make two points in this context:
\begin{itemize}
\item first, it is clear from Figure 12 than $\sim10\%$ of the known ESP host stars have 
velocities, V$_{tot}$, within a few kms$^{-1}$ of that of the Sun. Indeed, the transiting system, 
HD 209458, has a space motion with respect to the LSR which is almost identical with that of the Sun, 
while HD 114783 has a relative motion of only 6.2 kms$^{-1}$. Gonzalez {\sl et al.} (2001) derive
age estimates of 3 Gyrs (isochrones) and 4.3 Gyrs (activity) for the former star.
HD 114783 is too red to allow reliable 
an isochrone-based age estimate, but Vogt {\sl et al.} (2001) note that HD 114783
is chromospherically inactive, $\log{R'_{HK}}$=-4.96, or $\sim4.8$ Gyrs for the Donahue (1993) calibration.
Both stars are therefore likely to have ages similar to that of the Sun.
\item second, the Weak Anthropic Principle (WAP) can be expressed in two ways: as a positive concept, 
in that the planetary environment must permit the development of advanced lifeforms;
or as a less restrictive, negative concept, in that the environment should not be inimical
to the development of advanced lifeforms. Whether one chooses to express the WAP as a positive
or a negative concept depends on other issues, notably belief in the likelihood of life
developing elsewhere in the Universe. 
In either case, with a current sample of one known inhabited planet, the WAP should be given the 
same scientific weight as its converse, the Copernican Principle (``we're not special'').
Both are interesting philosophical concepts, which may have explanatory power; neither carries any 
evidentiary weight in the present context.
\end{itemize}

Finally, we have compared the distribution of properties of the extrasolar planetary systems against
the systemic velocities to search for possible trends or correlations. The only potentially
significant result is shown in the uppermost panel of Figure 13, plotting M$_2 \sin{i}$ against
velocity perpendicular to the Plane. The data suggest that, with the exception of HD 114762, higher-mass 
companions tend to be found in systems with low W velocities. The result is statistically marginal, 
but might indicate a correlation with the mass of the parent circumstellar disk. Clearly more
data are required to confirm whether this effect is real. 

\section {Summary and conclusions}

Over sixty stars with planetary-mass companions are now known. While these stars neither constitute 
a volume-limited sample nor, probably, a complete sampling of the full range of 
planetary systems, they provide sufficient numbers for a preliminary investigation of the
characteristics of the parent stars. In this paper we have compared the chemical abundance
distribution and kinematics of those stars against data for representative samples of the
local disk. Our metallicities are based on Str\"omgren photometry, using the calibration
derived by Schuster \& Nissen (1989). We have shown that the resulting metallicity scale
is offset to lower abundances with respect to recent high-resolution spectroscopic measurements.
This discrepancy is not important for our purposes, since Str\"omgren photometry is available
for 86\% of our reference sample - an Hipparcos-selected sample of 486 F, G and K stars within 
25 parsecs of the Sun. The abundance distributions derived for both datasets are therefore internally
consistent, although the solar abundance on this scale is $[Fe/H]_{uvby} \sim -0.1$ dex.

Comparing the abundance distributions of the two datasets, it is clear that, as
noted in previous studies, systems currently known to have extrasolar planets are
heavily weighted to high metallicities. We have used the fraction of known ESP
systems in the volume-complete sample (20/486, or 4.1\%) to set the two
distributions on a common scaling, and compute the observed frequency as a function of
chemical abundance. The results show a strong trend with abundance, with effectively 100\% frequency
at $[Fe/H]_{uvby} > 0.3$. However, even at abundances of less than $1 \over 2$ solar, 1 to 2\% of
stars are likely to have planetary-mass companions in the mass/semi-major axis/eccentricity range detectable
using current techniques. Clearly, these statistics represent a lower limit to the actual
frequency of extrasolar planetary systems. 

How rare are solar-abundance F, G, and K stars with planets? Note that while the
frequency of ESP hosts increases with [Fe/H], the absolute number of systems declines rapidly
at high abundances. Thus, planetary systems with parent stars of near-solar abundance contribute
a significant fraction of the total current sample.  Based on the full
FGK25 Hipparcos dataset, the local number density of stars with metallicities
within $\pm0.15$ dex of the solar abundance is 0.0044 stars pc$^{-3}$. The 
corresponding number density of ESP host stars, based on the data plotted in Figure 7, 
is 0.00018 stars pc$^{-3}$. Consider an annulus centred on the Solar Radius, $R_\odot = 8$ kpc., 
diameter 50 parsecs. Extrapolating from the local sample, we expect $\sim17,500$ ESP host
stars within this very limited subset of the Galactic Disk. Casting the net wider, consider
a wedge, thickness (perpendicular to the Plane) 50 parsecs, between Galactic 
radii of 7 and 9 kiloparsecs, a range
which encompasses relatively minor changes in mean abundance and stellar number density.
Based on our calculations, we would expect a total of over 900,000 solar-type stars
with Jovian-mass planetary companions. 

We have also compared the kinematics of the ESP host stars against the local Galactic disk via observations of a
volume-limited sample of M dwarfs. The planetary hosts exhibit a velocity distribution
which is relatively well matched to a Gaussian in each component, but with lower dispersions than in
the field-star sample. This suggests that the average age is only $\sim60\%$ that of a
representative subset of the disk. This may reflect the higher proportion of 
metal-rich stars in the ESP host sample. Individual stars, however, span a wide range
of motions, with velocities of up to 50-60 kms$^{-1}$ with respect to the Local Standard of Rest,
and no obvious correlation between kinematics and abundance. 

\acknowledgements 
I would like to thank Geoff Marcy for providing radial velocity measurements for
several stars in advance of publication; David Trilling, for useful comments; 
and David Koerner, for interesting discussion and sparking my initial interest in this topic.
The research for this paper made extensive use of the SIMBAD database, maintained by Strasberg Observatory,
and of Jean Schneider's `Extrasolar Planets Encyclopedia'.

\clearpage

\begin{deluxetable}{rrrrcrrrrl}
\tablewidth{0pt}
\tablecolumns{10}
\tablecaption{The host stars}
\tablehead{
\colhead{HIP}  & \colhead{Name} &  \colhead{M$_V$} & \colhead{(B-V)} & 
\colhead{$\pi$} & \colhead{M$_2\sin{i}$} & \colhead{a$\sin{i}$} & \colhead{P} & \colhead{e} & 
\colhead{Comments}  \\
    &      &       &       & \colhead{mas}   & \colhead{M$_J$}   & \colhead{AU}& 
\colhead{days}}  
\startdata
 \nodata &  BD -10 3166  &   5.5:&   0.84&   12$\pm$  4&   0.48&   0.05&    3.49&   0.00& 1   \\
     522 &  HD 142       &   3.65&   0.52&   39.00$\pm$  0.64&   1.36&   0.98&  338.00&   0.37&   \\
    1292 &  HD 1237      &   5.36&   0.75&   56.76$\pm$  0.53&   3.31&   0.49&  133.82&   0.51&    \\
    3479 &    HD 4208    &   5.21&   0.67&   30.58$\pm$  1.08&   0.80&   1.70&  829.00&   0.04&    \\
    3502 &    HD 4203    &   4.22&   0.73&   12.85$\pm$  1.27&   1.64&   1.09&  406.00&   0.53& 1, subgiant    \\
    5054 &  HD 6434      &   4.69&   0.60&   24.80$\pm$  0.89&   0.48&   0.15&   22.09&   0.30&    \\
    6643 &    HD 8574    &   3.90&   0.58&   22.65$\pm$  0.82&   2.23&   0.76&  228.80&   0.40&    \\
    7513 &  HD 9826      &   3.44&   0.54&   74.25$\pm$  0.72&   0.71&   0.06&    4.62&   0.03& $\upsilon$ And, 2    \\
    8159 &   HD 10697    &   3.73&   0.67&   30.71$\pm$  0.81&   6.59&   2.00& 1083.00&   0.12&  subgiant   \\
    9683 &   HD 12661    &   4.59&   0.72&   26.91$\pm$  0.83&   2.83&   0.78&  264.50&   0.33&    \\
   10138 &   HD 13445    &   5.98&   0.77&   91.63$\pm$  0.61&   4.00&   0.11&   15.78&   0.05&    \\
   12048 &   HD 16141    &   4.00&   0.71&   27.85$\pm$  1.39&   0.21&   0.35&   75.82&   0.28& subgiant    \\
   12653 &   HD 17051    &   4.22&   0.57&   58.00$\pm$  0.55&   2.26&   0.92&  320.10&   0.16& $\iota$ Hor   \\
   14954 &   HD 19994    &   3.31&   0.57&   44.69$\pm$  0.75&   2.00&   1.30&  454.00&   0.20&    \\
   16537 &  HD 22049     &   6.19&   0.88&  310.74$\pm$  0.85&   0.86&   3.30& 2502.10&   0.61& $\epsilon$ Eri   \\
   17096 & HD 23079      &   4.42&   0.58&   28.90$\pm$  0.56&   2.54&   1.48&  627.30&   0.06&   \\
   19921 &   HD 27442    &   3.14&   1.08&   54.84$\pm$  0.50&   1.43&   1.18&  423.00&   0.02&   subgiant  \\
   20723 &   HD 28185    &   4.82&   0.71&   25.28$\pm$  1.08&   5.60&   1.00&  385.00&   0.06&    \\
   24205 &   HD 33636    &   4.77&   0.58&   34.85$\pm$  1.33&   7.70&   2.60& 1553.00&   0.39&    \\
   26381 &   HD 37124    &   5.07&   0.67&   30.08$\pm$  1.15&   1.04&   0.58&  155.00&   0.19&    \\
   26394 &  HD 39091     &   4.35&   0.60&   54.92$\pm$  0.45&  10.37&   3.34& 2115.30&   0.62& \\
   27253 &  HD 38529     &   2.80&   0.74&   23.57$\pm$  0.92&   0.81&   0.13&   14.41&   0.28&  subgiant   \\
   31246 &   HD 46375    &   5.22&   0.86&   29.93$\pm$  1.07&   0.25&   0.04&    3.02&   0.02&    \\
   33212 &   HD 50554    &   4.40&   0.53&   32.23$\pm$  1.01&   4.90&   2.38& 1279.00&   0.42&    \\
   33719 &   HD 52265    &   4.06&   0.54&   35.63$\pm$  0.84&   1.13&   0.49&  118.96&   0.29&    \\
   40687 &   HD 68988    &   4.36&   0.62&   17.00$\pm$  0.96&   1.90&   0.07&    6.28&   0.14&    \\
   42723 &   HD 74156    &   3.57&   0.54&   15.49$\pm$  1.01&   1.56&   0.28&   51.61&   0.65& 3    \\
   43177 &   HD 75289    &   4.05&   0.58&   34.55$\pm$  0.56&   0.42&   0.05&    3.51&   0.00&    \\
   43587 &   HD 75732    &   5.46&   0.87&   79.80$\pm$  0.84&   0.84&   0.11&   14.65&   0.05& 55 Cnc   \\
   45982 &   HD 80606    &   5.10&   0.72&   17.13$\pm$  5.77&   3.41&   0.44&  111.78&   0.93&    \\
   47007 &   HD 82943    &   4.35&   0.59&   36.42$\pm$  0.84&   0.88&   0.73&  221.60&   0.54&  4  \\
   47202 &  HD 83443     &   5.05&   0.79&   22.97$\pm$  0.90&   0.35&   0.04&    2.99&   0.08&  5  \\
   50786 &   HD 89744    &   2.79&   0.49&   25.65$\pm$  0.70&   7.20&   0.88&  256.00&   0.70&    \\
   52409 &   HD 92788    &   4.76&   0.69&   30.94$\pm$  0.99&   3.80&   0.94&  340.00&   0.36&    \\
   53721 &   HD 95128    &   4.36&   0.56&   71.04$\pm$  0.66&   2.41&   2.10& 1096.00&   0.10& 47 UMa, 6   \\
   59610 &  HD 106252    &   4.49&   0.64&   26.71$\pm$  0.91&   6.81&   2.61& 1500.00&   0.54&    \\
   60644 &  HD 108147    &   4.07&   0.50&   25.93$\pm$  0.69&   0.34&   0.10&   10.88&   0.56&    \\
   64426 & HD 114762     &   4.26&   0.52&   24.65$\pm$  1.44&  11.00&   0.30&   84.03&   0.33&  7 \\
   64457 &  HD 114783    &   6.02&   0.91&   48.95$\pm$  1.06&   0.99&   1.20&  501.00&   0.10&    \\
   65721 &  HD 117176    &   3.71&   0.69&   55.22$\pm$  0.73&   6.60&   0.43&  116.60&   0.40& 70 Vir,  subgiant   \\
   67275 &  HD 120136    &   3.53&   0.48&   64.12$\pm$  0.70&   3.87&   0.05&    3.31&   0.02& $\tau$ Boo   \\
   68162 &  HD 121504    &   4.30&   0.59&   22.54$\pm$  0.91&   0.89&   0.32&   64.60&   0.13&    \\
   72339 &  HD 130332    &   5.68&   0.75&   33.60$\pm$  1.51&   1.08&   0.09&   10.72&   0.05&    \\
   74500 &  HD 134987    &   4.40&   0.70&   38.98$\pm$  0.98&   1.58&   0.78&  260.00&   0.25&    \\
   77740 &  HD 141937    &   4.63&   0.60&   29.89$\pm$  1.08&   9.70&   1.49&  658.80&   0.40&    \\
   78459 &  HD 143761    &   4.19&   0.61&   57.38$\pm$  0.71&   1.10&   0.23&   39.47&   0.03& $\rho$ CrB   \\
   79248 &  HD 145675    &   5.38&   0.90&   55.11$\pm$  0.59&   3.30&   2.50& 1619.00&   0.35& 14 Her   \\
   86796 &  HD 160691    &   4.23&   0.70&   65.46$\pm$  0.80&   1.97&   1.65&  743.00&   0.62& subgiant    \\
   87330 &  HD 162020    &   6.63&   0.96&   31.99$\pm$  1.48&  13.73&   0.08&    8.43&   0.28&    \\
   89844 &  HD 168443    &   4.03&   0.70&   26.40$\pm$  0.85&   7.20&   0.29&   57.90&   0.54&  8, subgiant   \\
   90004 &  HD 168746    &   4.78&   0.69&   23.19$\pm$  0.96&   0.24&   0.07&    6.41&   0.00&    \\
   90485 &  HD 169830    &   3.11&   0.48&   27.53$\pm$  0.91&   2.96&   0.82&  230.40&   0.34&    \\
   93746 &  HD 177830    &   3.32&   1.09&   16.94$\pm$  0.76&   1.28&   1.00&  391.00&   0.43&  subgiant   \\
   94076 &  HD 178911    &   3.29&   0.63&   20.42$\pm$  1.57&   6.29&   0.33&   71.49&   0.12&  subgiant   \\
   94645 &  HD 179949    &   4.09&   0.51&   36.97$\pm$  0.80&   0.84&   0.05&    3.09&   0.05&    \\
   96901 &  HD 186427    &   4.55&   0.66&   46.70$\pm$  0.52&   1.50&   1.70&  804.00&   0.67&  16 CygB  \\
   97336 &  HD 187123    &   4.46&   0.61&   20.87$\pm$  0.71&   0.52&   0.04&    3.10&   0.03&    \\
   98714 &  HD 190228    &   3.34&   0.75&   16.10$\pm$  0.81&   4.99&   2.31& 1127.00&   0.43&  subgiant   \\
   99711 &  HD 192263    &   6.30&   0.94&   50.27$\pm$  1.23&   0.76&   0.15&   23.87&   0.03&    \\
  100970 &  HD 195019    &   4.05&   0.64&   26.77$\pm$  0.89&   3.43&   0.14&   18.30&   0.05&    \\
  104903 &  HD 202206    &   4.75&   0.71&   21.58$\pm$  1.14&  14.68&   0.80&  258.97&   0.42&    \\
  108859 &  HD 209458    &   4.29&   0.53&   21.24$\pm$  1.00&   0.69&   0.05&    3.53&   0.00&    \\
  109378 &  HD 210277    &   4.90&   0.77&   46.97$\pm$  0.79&   1.28&   1.10&  437.00&   0.45&    \\ 
  111143 &  HD 213240    &   3.75&   0.61&   24.54$\pm$  0.81&   3.70&   1.60&  759.00&   0.31&    \\
  113020 &  Gl 876       &  11.81&   1.60&  212.69$\pm$  2.10&   1.98&   0.21&   61.02&   0.27& 9   \\
  113357 &  HD 217014    &   4.56&   0.66&   65.10$\pm$  0.76&   0.47&   0.05&    4.23&   0.00& 51 Peg    \\
  113421 &  HD 217107    &   4.71&   0.72&   50.71$\pm$  0.75&   1.28&   0.07&    7.11&   0.14&    \\
  116906 &  HD 222582    &   4.59&   0.60&   23.84$\pm$  1.11&   5.40&   1.35&  576.00&   0.71&    \\
\enddata
\tablecomments{Column 1 lists the Hipparcos designation, where appropriate; column 2 gives
the common name; columns 3 and 4 list M$_V$ and (B-V), generally derived from the literature data
listed in the Hipparcos catalogue; column 5 lists the parallax and associated uncertainty (in
milliarcseconds); column 6, 7, 8 and 9 list the mass, semi-major axis, period and eccentricity
of the planetary-mass companions. The latter data are taken from J. Schneider's on-line
catalogue ({\sl http://cfa-www.harvard.edu/planets/catalog.html}), except for HD 162020 and HD 202206, 
where the data are from the Geneva Observatory web-site ({\sl http://obswww.unige.ch/~udry/planet/}). 
Named stars, evolved stars and stars
with more than one companion (see below) are identified in the final column. \\
Notes for individual stars: \\
1. Selected for monitoring based on high metallicity (Butler {\sl et al.}, 2000) \\
2. At least 2 other planetary-mass companions, M$ \sin{i} = 2.11, 4.61 M_J$ at a$\sin{i}$ = 0.83, 2.50 AU \\
3. At least 1 other planet/brown dwarf companion, M$ \sin{i} = 7.5 M_J$ at a$\sin{i}$ = 4.47 AU \\
4. At least 1 other planet companion, M$ \sin{i} = 1.63 M_J$ at a$\sin{i}$ = 1.16 AU \\
5. At least 1 other planet companion, M$ \sin{i} = 0.16 M_J$ at a$\sin{i}$ = 0.174 AU \\
6. At least 1 other planet companion, M$ \sin{i} = 0.76 M_J$ at a$\sin{i}$ = 3.73 AU \\
7. Suspected high inclination, implying brown dwarf companion \\
8. At least 1 other planet/brown dwarf companion, M$ \sin{i} = 17.1 M_J$ at a$\sin{i}$ = 2.87 AU \\
9. At least 1 other planet companion, M$ \sin{i} = 0.56 M_J$ at a$\sin{i}$ = 0.13 AU }
\end{deluxetable}
\clearpage

\begin{deluxetable}{rrrrrrrr}
\tablewidth{0pt}
\tablecolumns{8}
\tablecaption{Str\"omgren data and abundance measurements}
\tablehead{
\colhead{HIP} & \colhead{[Fe/H]$_{sp}$} & \colhead {ref.} &
\colhead{(b-y)} & \colhead {m$_1$} &
\colhead{c$_1$} & \colhead{[Fe/H]$_{uvby}$}&  \colhead{$\delta$[Fe/H]} }
\startdata
    522&   0.04&   10&   0.332&   0.168&   0.416&  -0.06&  -0.10 \\
    1292&   0.11&    1&   0.459&   0.289&   0.300&  -0.06&  -0.17 \\
    3479&  -0.24&    9&   0.413&   0.213&   0.285&  -0.22&   0.02 \\
    3502&   0.22&    9&   0.467&   0.288&   0.392&   0.22&   0.00 \\
    5054&  -0.55&    1&   0.384&   0.159&   0.274&  -0.54&   0.01 \\
    6643&  -0.09&   11&   0.362&   0.169&   0.378&  -0.22&  -0.13 \\
    7513&   0.12&    2&   0.346&   0.176&   0.415&  -0.02&  -0.14 \\
    8159&   0.16&    3&   0.440&   0.238&   0.379&   0.08&  -0.08 \\
    9683&   0.35&    3&   0.448&   0.267&   0.398&   0.25&  -0.10 \\
   10138&  -0.20&    1&   0.484&   0.337&   0.287&  -0.14&   0.06 \\
   12048&   0.15&    1&   0.422&   0.213&   0.378&  -0.02&  -0.17 \\
   12653&   0.25&    1&   0.357&   0.188&   0.364&   0.08&  -0.17 \\
   14594&   0.26&    1&   0.361&   0.185&   0.422&   0.00&  -0.26 \\
   16537&  -0.07&    1&   0.504&   0.430&   0.263&  -0.28&  -0.21 \\
   17096&\nodata&     &   0.369&   0.179&   0.330&  -0.14&    \\
   19921&   0.22&   12&   0.651&   0.513&   0.406&   0.26&   0.04 \\
   20723&   0.24&    1&   0.443&   0.264&   0.352&   0.15&  -0.09 \\
   24205&  -0.13&    9&   0.378&   0.177&   0.324&  -0.20&  -0.07 \\
   26381&  -0.41&    3&   0.421&   0.202&   0.280&  -0.37&   0.04 \\
   26394&\nodata&     &   0.371&   0.193&   0.363&   0.09&   \\
   27253&   0.39&    1&   0.471&   0.278&   0.437&   0.23&  -0.16 \\
   31246&   0.21&    3&   0.502&   0.401&   0.337&   0.01&  -0.20 \\
   33212&   0.02&   11&   0.366&   0.179&   0.347&  -0.12&  -0.14 \\
   33719&   0.24&    1&   0.360&   0.190&   0.404&   0.08&  -0.16 \\
   40687&   0.24&    9&   0.405&   0.244&   0.387&   0.36&   0.12 \\
   42723&   0.13&   11&   0.375&   0.181&   0.390&  -0.06&  -0.19 \\
   43177&   0.27&    1&   0.360&   0.191&   0.405&   0.10&  -0.17 \\
   43587&   0.45&    4&   0.536&   0.357&   0.415&   0.10&  -0.35 \\
   45982&   0.43&    5&   0.470&   0.312&   0.361&   0.20&  -0.23 \\
   47007&   0.33&    1&   0.386&   0.217&   0.390&   0.27&  -0.06 \\
   47202&   0.39&    1&   0.488&   0.349&   0.368&   0.18&  -0.21 \\
   50786&   0.30&    3&   0.338&   0.184&   0.451&   0.14&  -0.16 \\
   52409&   0.31&    3&   0.433&   0.253&   0.376&   0.22&  -0.09 \\
   53721&   0.01&    6&   0.391&   0.202&   0.343&   0.02&   0.01 \\
   59610&  -0.16&   11&   0.390&   0.187&   0.341&  -0.13&   0.03 \\
   60644&   0.20&    1&   0.346&   0.177&   0.391&  -0.01&  -0.21 \\
   64426&  -0.60&    6&   0.365&   0.125&   0.297&  -0.74&  -0.14 \\
   64457&   0.33&    9&   0.521&   0.458&   0.309&  -0.17&  -0.50 \\
   65721&  -0.01&    6&   0.446&   0.232&   0.351&  -0.07&  -0.06 \\
   67275&   0.32&    7&   0.318&   0.177&   0.439&   0.13&  -0.19 \\
   68162&   0.17&    1&   0.381&   0.189&   0.361&  -0.02&  -0.19 \\
   72339&   0.05&    3&   0.475&   0.305&   0.316&  -0.02&  -0.07 \\
   74500&   0.32&    3&   0.435&   0.256&   0.374&   0.22&  -0.10 \\
   77740&   0.16&   11&   0.388&   0.225&   0.346&   0.24&   0.08 \\
   78459&  -0.29&    6&   0.394&   0.178&   0.337&  -0.27&   0.02 \\
   79248&   0.50&    8&   0.537&   0.366&   0.438&   0.13&  -0.37 \\
   86796&   0.28&   10&   0.432&   0.244&   0.393&   0.20&  -0.08 \\
   87330&   0.01&    1&   0.579&   0.534&   0.244&   0.11&   0.10 \\
   89844&   0.10&    3&   0.455&   0.233&   0.377&  -0.06&  -0.16 \\
   90004&  -0.06&    1&   0.435&   0.223&   0.342&  -0.09&  -0.03 \\
   90485&   0.22&    1&   0.328&   0.177&   0.446&   0.09&  -0.13 \\
   94076&   0.28&   13&   0.403&   0.219&   0.378&   0.16&  -0.12 \\
   94645&   0.22&   14&   0.346&   0.183&   0.384&   0.08&  -0.14 \\
   96901&   0.07&    7&   0.416&   0.226&   0.354&   0.09&   0.02 \\
   97336&   0.16&    8&   0.405&   0.224&   0.365&   0.17&   0.01 \\
   98714&  -0.24&    1&   0.482&   0.264&   0.306&  -0.27&  -0.03 \\
   99711&  -0.03&    3&   0.541&   0.493&   0.275&  -0.20&  -0.17 \\
  100970&   0.16&   15&   0.419&   0.204&   0.362&  -0.11&  -0.27 \\
  104903&   0.37&    1&   0.435&   0.253&   0.390&   0.24&  -0.13 \\
  108859&   0.04&    3&   0.361&   0.174&   0.362&  -0.15&  -0.19 \\
  109378&   0.23&    1&   0.466&   0.285&   0.369&   0.16&  -0.07 \\
  111143&   0.16&    1&   0.387&   0.190&   0.399&  -0.02&  -0.18 \\
  113357&   0.21&    3&   0.416&   0.233&   0.371&   0.18&  -0.03 \\
  113421&   0.39&    1&   0.456&   0.299&   0.376&   0.28&  -0.11 \\
  116906&  -0.01&   16&   0.406&   0.202&   0.345&  -0.08&  -0.07 \\

\enddata
\tablecomments{ Column 2 lists abundances derived from high-resolution
spectroscopy from the following sources: \\
1. Santos {\sl et al.}, 2001; 2. Gonzalez \& Laws, 2000; 3. Gonzalez 
{\sl et al.}, 2001; 4. Gonzalez \& Vanture, 1998; 5. Naef {\sl et al.}, 2001;
6. Gonzalez, 1998; 7. Laws \& Gonzalez, 2001; 8. Gonzalez {\sl et al.}, 1999;
9. Vogt {\sl et al.}, 2001a; 10. Favata {\sl et al.}, 1997; 
11. Geneva Observatory ({\sl http://obswww.unige.ch/~udry/planet/});
12. Randich {\sl et al.}, 1999; 13. Zucker {\sl et al.}, 2001;
14. Tinney {\sl et al.}, 2001; 15. Santos {\sl et al.}, 2000; 
16. Vogt {\sl et al.}, 2001b. \\
Columns 4, 5 and 6 list Str\"omgren (b-y), m$_1$ and c$_1$ data from
Hauck \& Mermilliod (1998), and 
Column 7 lists [Fe/H] derived from those data using the Sch\"uster \&
Nissen (1989) calibration. \\
Column 8 lists $\delta$[Fe/H] = [Fe/H]$_{uvby}$ - [Fe/H]$_{sp}$. } 
\end{deluxetable}
\clearpage

\begin{deluxetable}{rrrrcrrrrrrl}
\tablewidth{0pt}
\tablecolumns{11}
\tablecaption{Space motions}
\tablehead{
\colhead{HIP}  & \colhead{$\pi$} &  \colhead{$\mu_\alpha$} & \colhead{$\mu_\delta$} & 
\colhead{V$_{rad}$} & \colhead{ref.} & \colhead{U} & \colhead{V} & \colhead{W} & \colhead{V$_{LSR}$} &
\colhead{Comments}  \\
 \colhead{}   &  \colhead{mas}     &  \colhead{mas}      &  \colhead{mas}    & \colhead{kms$^{-1}$}   & \colhead {}
  & \colhead{kms$^{-1}$}& \colhead{kms$^{-1}$}& \colhead{kms$^{-1}$}& \colhead{kms$^{-1}$}}  
\startdata
     522&   39.00 $\pm$ 0.64&   575.2&   -39.9&     2.6&    2&   -58.2&   -37.2&   -12.1&    58.0&           \\
    1292&   56.76 $\pm$ 0.53&   433.9&   -57.9&    -5.8&    1&   -33.0&   -16.6&     2.7&    27.5&           \\
    3479&   30.58 $\pm$ 1.08&   313.5&   150.0&    55.4&    2&   -53.1&    -5.2&   -55.9&    65.0&           \\
    3502&   12.85 $\pm$ 1.27&   125.2&  -124.0&   -14.1&    7&   -16.5&   -59.3&   -25.2&    57.3&           \\
    5054&   24.80 $\pm$ 0.89&  -169.0&  -527.7&    23.0&    1&    85.4&   -66.6&    -3.4&   113.5&           \\
    6643&   22.65 $\pm$ 0.82&   252.6&  -158.6&    18.9&    1&   -44.3&   -37.0&   -30.3&    52.1&           \\
    7513&   74.25 $\pm$ 0.72&  -172.6&  -381.0&   -28.3&    2&    28.5&   -22.1&   -14.6&    42.7&           \\
    8159&   30.71 $\pm$ 0.81&   -45.0&  -105.4&   -43.5&    1&    35.6&   -26.7&    15.0&    55.0&           \\
    9683&   26.91 $\pm$ 0.83&  -107.8&  -175.3&   -52.2&    6&    55.1&   -31.7&    -0.1&    70.6&           \\
   10138&   91.63 $\pm$ 0.61&  2092.8&   654.5&    56.6&    1&   -97.5&   -75.9&   -28.5&   114.4&           \\
   12048&   27.85 $\pm$ 1.39&  -156.9&  -437.1&   -53.0&    2&    85.8&   -41.0&     2.4&   102.7&           \\
   12653&   58.00 $\pm$ 0.55&   333.7&   219.2&    15.5&    2&   -31.2&   -16.7&    -7.3&    24.1&           \\
   14594&   44.69 $\pm$ 0.75&  -209.6&   -69.2&    18.3&    2&     2.9&     9.0&   -28.2&    28.5&           \\
   16537&  310.74 $\pm$ 0.85&  -976.4&    18.0&    15.5&    2&    -3.3&     7.2&   -20.0&    19.1&           \\
   17096&   28.90 $\pm$ 0.56&  -193.6&   -91.9&   -22.2&    8&    29.1&    29.6&     1.4&    53.2&           \\
   19921&   54.84 $\pm$ 0.50&   -48.0&  -167.8&    29.3&    2&    15.1&   -22.1&   -19.3&    32.5&           \\
   20723&   25.28 $\pm$ 1.08&    80.8&   -31.1&    50.3&    1&   -49.2&   -15.2&   -11.7&    40.7&           \\
   24205&   34.85 $\pm$ 1.33&   180.8&  -137.3&    -1.0&    6&     5.8&   -28.2&    11.1&    33.3&           \\
   26381&   30.08 $\pm$ 1.15&   -79.8&  -420.0&   -12.0&    2&    21.6&   -47.4&   -44.4&    64.4&           \\
   26394&   54.92 $\pm$ 0.45&   312.0&  1050.2&     9.4&    2&   -82.9&   -46.4&     0.5&    84.1&           \\
   27253&   23.57 $\pm$ 0.92&   -80.1&  -141.8&    28.9&    2&   -12.6&   -24.8&   -33.7&    33.0&           \\
   31246&   29.93 $\pm$ 1.07&   114.2&   -96.8&     4.0&    2&     6.0&   -21.5&     8.8&    27.9&           \\
   33212&   32.23 $\pm$ 1.01&   -37.3&   -96.4&    -3.9&    1&     3.6&   -10.0&   -11.5&    15.0&           \\
   33719&   35.63 $\pm$ 0.84&  -115.8&    80.3&    53.8&    1&   -52.2&   -21.0&    -8.7&    45.1&           \\
   40687&   17.00 $\pm$ 0.96&   128.3&    31.7&   -69.5&    7&    75.1&   -21.0&   -10.6&    86.6&           \\
   42723&   15.49 $\pm$ 1.01&    25.0&  -200.5&     3.8&    1&    28.8&   -51.7&   -18.4&    61.5&           \\
   43177&   34.55 $\pm$ 0.56&   -20.5&  -227.7&     9.3&    1&    20.9&   -12.5&   -21.8&    34.9&           \\
   43587&   79.80 $\pm$ 0.84&  -485.5&  -234.4&    26.6&    2&   -36.5&   -18.2&    -8.1&    29.5&           \\
   45982&   17.13 $\pm$ 5.77&    47.0&     6.9&     3.8&    1&     6.9&     2.9&    11.4&    26.5&           \\
   47007&   36.42 $\pm$ 0.84&     2.4&  -174.1&     8.1&    1&    10.3&   -19.8&    -8.9&    25.0&           \\
   47202&   22.97 $\pm$ 0.90&    22.4&  -120.8&    28.8&    1&    20.0&   -30.4&   -12.1&    39.4&           \\
   50786&   25.65 $\pm$ 0.70&  -120.2&  -138.6&    -6.5&    2&   -10.5&   -29.7&   -14.1&    25.4&           \\
   52409&   30.94 $\pm$ 0.99&   -12.6&  -222.8&    -4.5&    1&    16.1&   -22.2&   -20.9&    34.0&           \\
   53721&   71.04 $\pm$ 0.66&  -315.9&    55.2&    12.6&    2&   -24.6&    -2.6&     2.1&    17.5&           \\
   59610&   26.71 $\pm$ 0.91&    23.8&  -279.4&    15.5&    1&    28.8&   -43.4&     0.3&    54.9&           \\
   60644&   25.93 $\pm$ 0.69&  -181.6&   -60.8&    -5.1&    1&   -30.4&   -11.6&   -13.9&    22.4&           \\
   64426&   24.65 $\pm$ 1.44&  -582.7&    -2.0&    49.9&    2&   -81.8&   -69.9&    59.0&   117.0&           \\
   64457&   48.95 $\pm$ 1.06&  -138.1&     9.6&   -12.0&    7&   -15.5&    -2.8&    -8.7&     6.2&           \\
   65721&   55.22 $\pm$ 0.73&  -234.8&  -576.2&     4.9&    2&    13.3&   -51.8&    -4.0&    52.1&           \\
   67275&   64.12 $\pm$ 0.70&  -480.3&    54.2&   -15.6&    2&   -33.5&   -19.0&    -6.3&    27.2&           \\
   68162&   22.54 $\pm$ 0.91&  -250.6&   -84.0&    19.5&    1&   -27.6&   -52.0&    -1.4&    50.3&           \\
   72339&   33.60 $\pm$ 1.51&  -129.6&  -140.8&   -12.5&    1&    -9.3&   -26.2&   -10.7&    21.2&           \\
   74500&   38.98 $\pm$ 0.98&  -399.0&   -75.1&     3.4&    2&   -21.6&   -39.6&    20.3&    45.5&           \\
   77740&   29.89 $\pm$ 1.08&    97.1&    24.0&    -3.0&    1&     2.8&    13.3&    -8.7&    22.6&           \\
   78459&   57.38 $\pm$ 0.71&  -196.9&  -773.0&    18.4&    2&    54.7&   -35.4&    20.8&    76.7&           \\
   79248&   55.11 $\pm$ 0.59&   132.5&  -298.4&   -13.9&    1&    23.8&   -12.2&   -16.3&    35.6&           \\
   86796&   65.46 $\pm$ 0.80&   -15.1&  -191.2&     9.0&    2&     2.9&   -14.4&    -7.6&    15.8&           \\
   87330&   31.99 $\pm$ 1.48&    21.0&   -25.2&   -27.5&    1&   -27.7&     2.7&    -1.0&    20.4&           \\
   89844&   26.40 $\pm$ 0.85&   -92.2&  -224.2&   -48.7&    1&   -29.7&   -57.9&    -6.1&    56.2&           \\
   90004&   23.19 $\pm$ 0.96&   -22.1&   -69.2&   -25.6&    1&   -19.4&   -22.3&    -2.9&    19.8&           \\
   90485&   27.53 $\pm$ 0.91&    -0.8&    15.2&   -17.2&    1&   -16.9&     1.1&     4.0&    14.6&           \\
   93746&   16.94 $\pm$ 0.76&   -40.7&   -51.8&   -74.0&    2&   -23.8&   -72.1&    -7.0&    68.2&           \\
   94076&   20.42 $\pm$ 1.57&    47.1&   194.5&   -40.4&    1&   -58.3&   -19.7&     1.4&    51.2&           \\
   94645&   36.97 $\pm$ 0.80&   114.8&  -101.8&   -24.7&    7&   -26.6&   -12.9&   -11.0&    18.7&           \\
   96901&   46.70 $\pm$ 0.52&  -135.1&  -163.5&   -27.1&    2&    17.8&   -29.6&    -1.7&    37.4&           \\
   97336&   20.87 $\pm$ 0.71&   143.1&  -123.2&   -17.5&    3&     2.3&   -16.0&   -43.4&    39.7&           \\
   98714&   16.10 $\pm$ 0.81&   104.9&   -69.8&   -50.2&    1&   -20.0&   -47.3&   -35.6&    51.6&           \\
   99711&   50.27 $\pm$ 1.23&   -63.4&   262.3&   -10.8&    1&   -16.4&    10.1&    19.8&    31.8&           \\
  100970&   26.77 $\pm$ 0.89&   349.5&   -56.9&   -92.7&    2&   -72.3&   -77.3&   -36.5&    99.6&           \\
  104903&   21.58 $\pm$ 1.14&   -38.2&  -119.8&    14.6&    1&    22.5&   -19.2&   -10.0&    35.5&           \\
  108859&   21.24 $\pm$ 1.00&    28.9&   -18.4&   -14.8&    1&    -5.6&   -15.6&     0.6&    13.7&           \\
  109378&   46.97 $\pm$ 0.79&    85.5&  -449.8&   -20.9&    3&     4.3&   -50.2&    -6.2&    47.1&           \\
  111143&   24.54 $\pm$ 0.81&  -135.2&  -194.1&    -0.5&    1&    25.6&   -29.9&    23.1&    52.9&           \\
  113020&  212.69 $\pm$ 2.10&   960.3&  -675.6&    -1.8&    5&   -12.5&   -20.0&   -11.5&    15.5&           \\
  113357&   65.10 $\pm$ 0.76&   208.1&    61.0&   -31.2&    2&   -14.9&   -28.0&    14.7&    31.9&           \\
  113421&   50.71 $\pm$ 0.75&    -6.1&   -16.0&   -12.1&    4&    -1.1&    -7.8&     9.3&    19.0&           \\
  116906&   23.84 $\pm$ 1.11&  -145.4&  -111.1&    12.1&    7&    36.6&    -0.3&   -11.5&    46.4&           \\
\enddata
\tablecomments{Columns 2, 3 and 4 list astrometric data from the Hipparcos catalogue; 
column 5 shows the measured radial velocity and column 6 gives the source, as follows: \\
1. Geneva Observatory  ({\sl http://obswww.unige.ch/~udry/planet/}) - measurements accurate to 1-10 ms$^{-1}$ \\
2. the compilation by Duflot {\sl et al.}, 1995 - measurements accurate to 2-5 kms$^{-1}$ \\
3. Marcy, as cited in Gonzalez (1999) \\
4. Griffin, 1972 - accurate to 0.4 kms$^{-1}$ \\
5. Marcy \& Benitz, 1989 - accurate to 0.3 kms$^{-1}$ \\
6. Fouts \& Sandage, 1986 \\
7. Nidever {\sl et al.} (2002) \\
8. Marcy (2001), priv. comm. \\
Columns 7, 8 and 9 list the space motions, and column 10 gives the velocity with respect to
the Local Standard of Rest.}

\end{deluxetable}

\clearpage
\begin{figure}
\figurenum{1}
\plotone{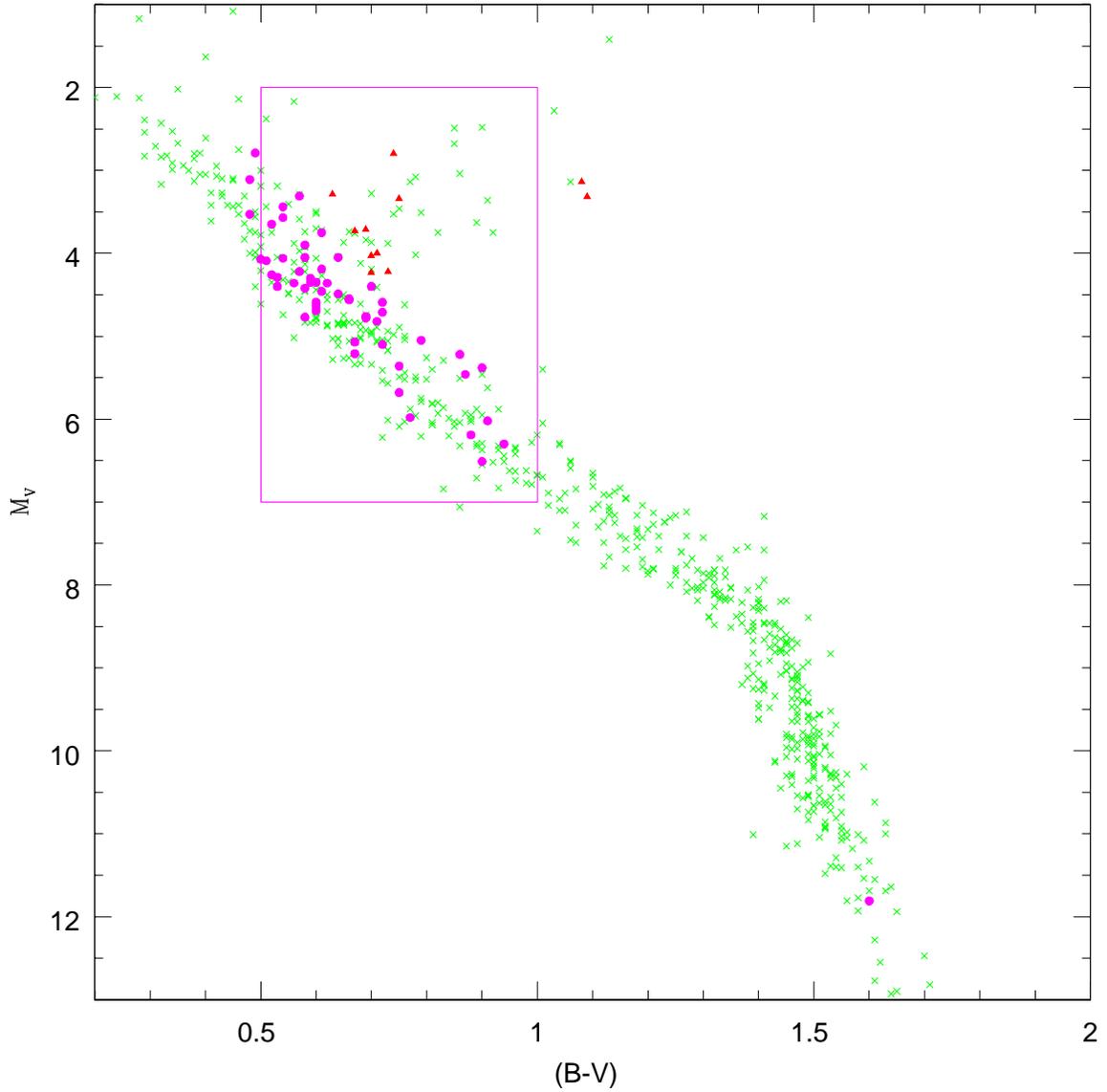}
\caption{The location of the host stars of planets in the (M$_V$, (B-V)) diagram.
The main-sequence is defined by stars with trigonometric parallaxes measured to an
accuracy of better than 10\%; main-sequence planetary hosts are plotted as solid points, 
evolved stars are plotted as solid triangles. The box marks the location of the 
reference sample (see Figure 4).}
\label{hrd}
\end{figure}

\begin{figure}
\figurenum{2}
\plotone{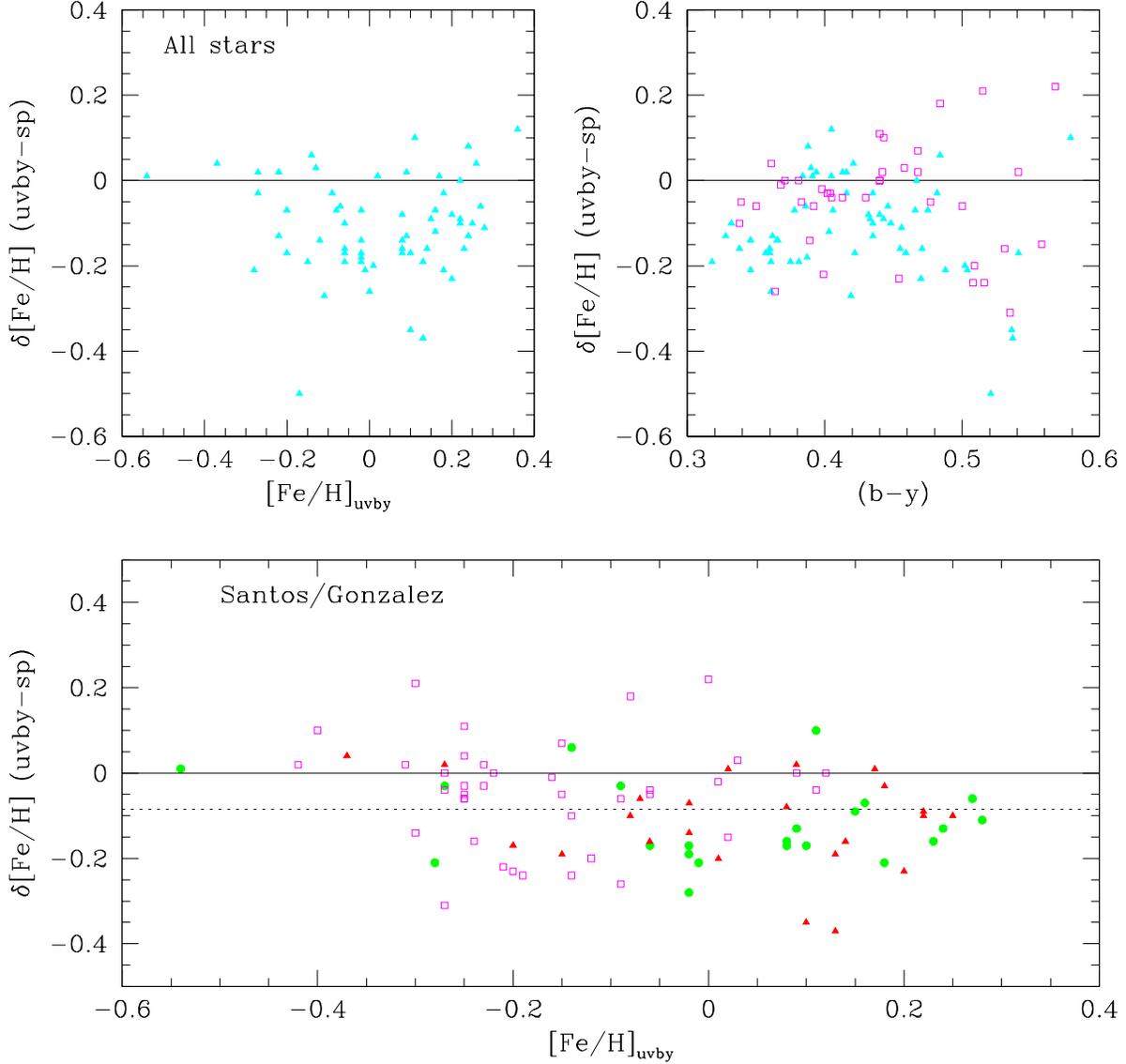}
\caption{Comparison between spectroscopic and Str\"omgren-based abundance
estimates for stars known to have planets. The upper two panels plot the
data given in Table 2 as solid triangles, with field stars from Santos {\sl et al.}
(2001) plotted as open squares in the upper right panel.
The lower panel plots data from Santos {\sl et al.} (2001:
solid points) and from the various analyses by Gonzalez and collaborators (solid
triangles). Data for field stars analysed by Santos {\sl et al.} are
plotted as open squqres. The dotted line marks the mean offset between these
abundance calibrations.}   
\label{fehcomp}
\end{figure}

\begin{figure}
\figurenum{3}
\plotone{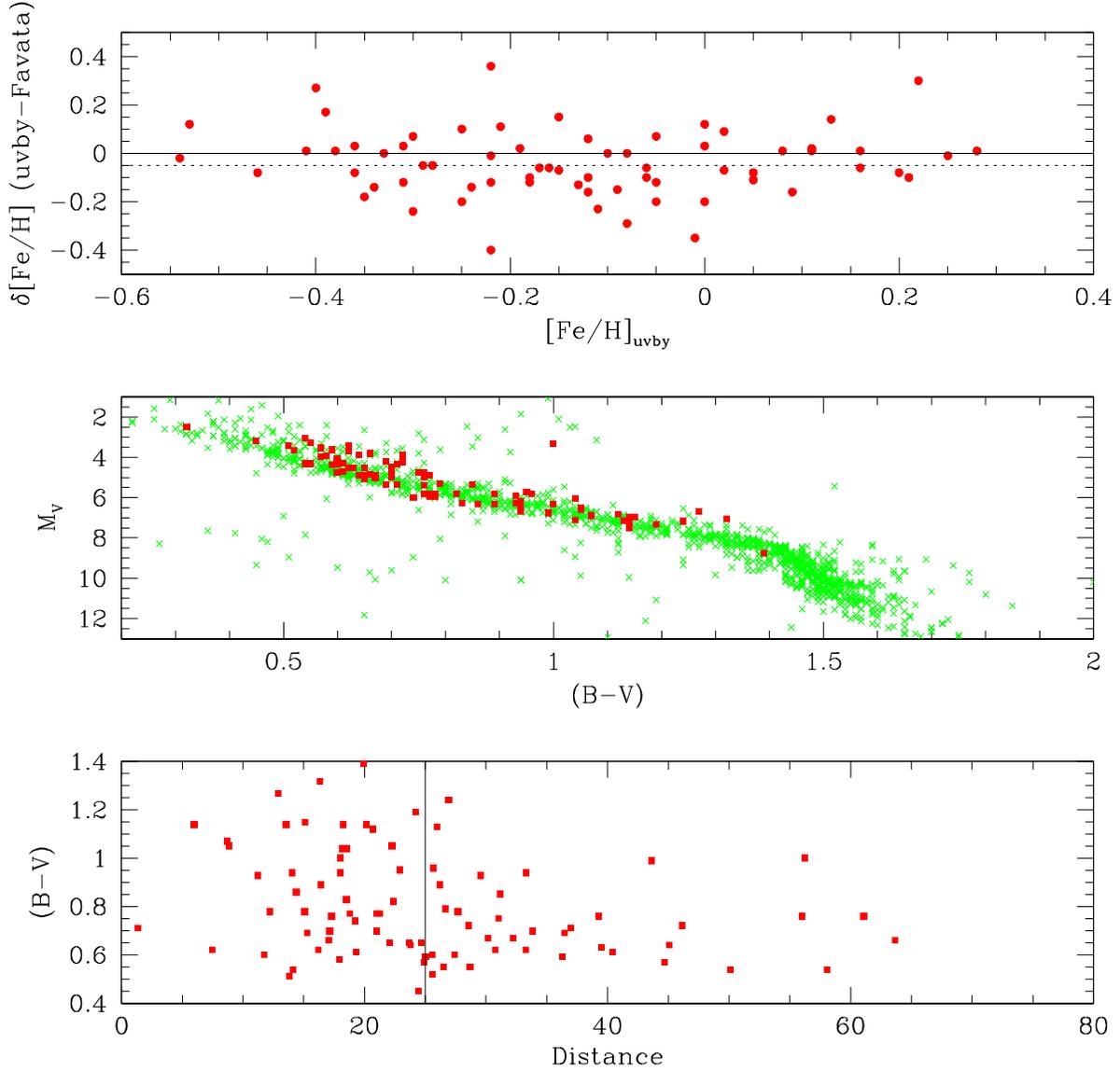}
\caption{Data for stars in the Favata {\sl et al.} (1997) disk dwarf sample. The
upper panel compares the abundances against the Str\"omgren calibration, with the
dotted line marking the mean offset. The middle panel shows the distribution in the HR 
diagram, where the solid squares mark the F97 stars. The lower panel plots
the distance distribution, with the vertical line marking the nominal distance
limit of the Gliese catalogue.}
\label{favata}
\end{figure}

\begin{figure}
\figurenum{4}
\plotone{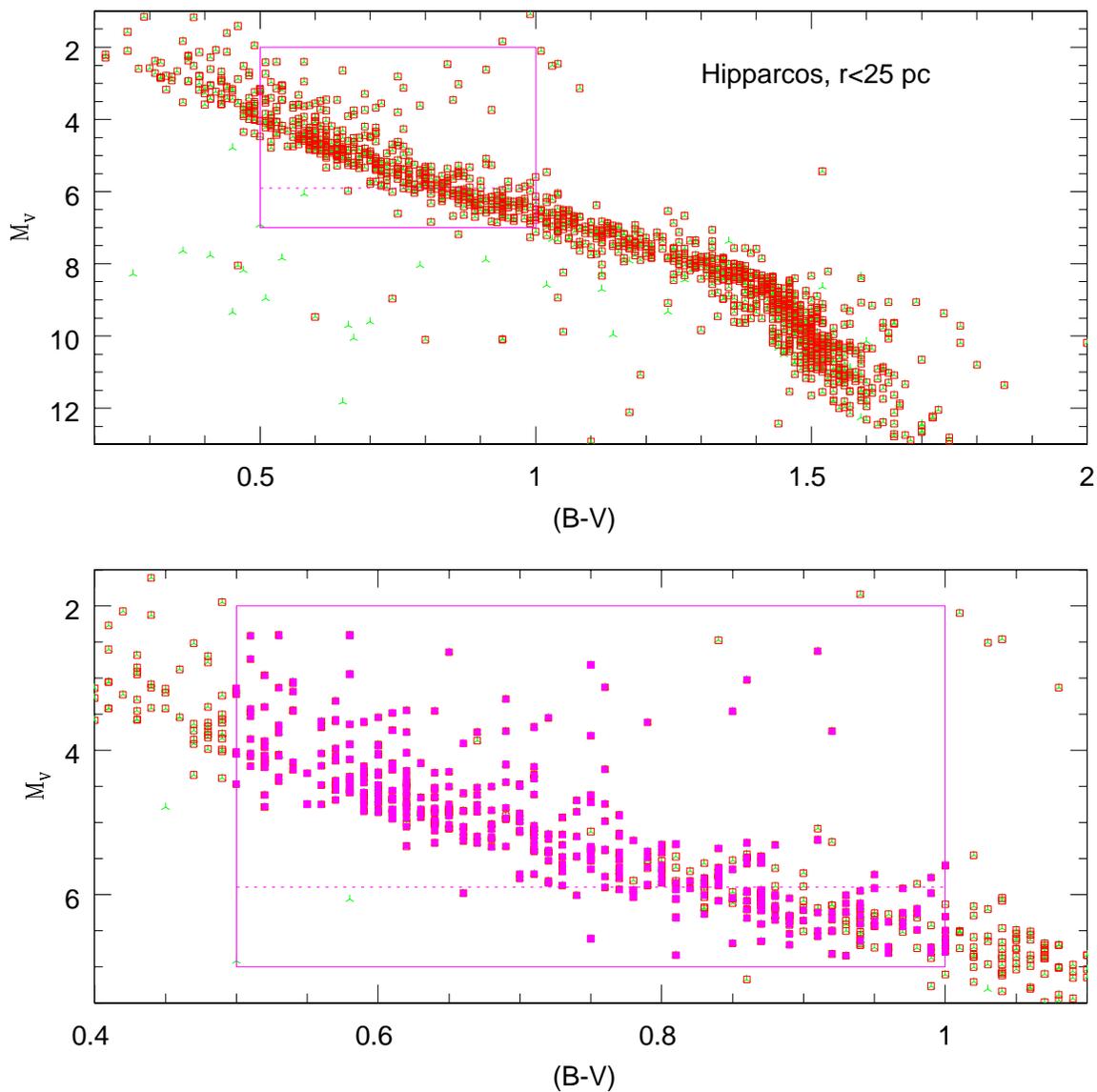}
\caption{ The (M$_V$, (B-V)) colour-magnitude diagram defined by 1549 stars
in the Hipparcos catalogue with $\pi > 40$ milliarcseconds. Open squares mark
1477 stars with ${\sigma_\pi \over \pi} < 0.2$; 
almost all of the stars lying below the main-sequence have
low-accuracy parallax measurements. The box outlines the limits of the FGK25 sample and
the dotted line shows the all-sky completeness limit. 
Solid squares in the lower panel mark stars with Str\"omgren photometry.  }
\label{plothip}
\end{figure}

\begin{figure}
\figurenum{5}
\plotone{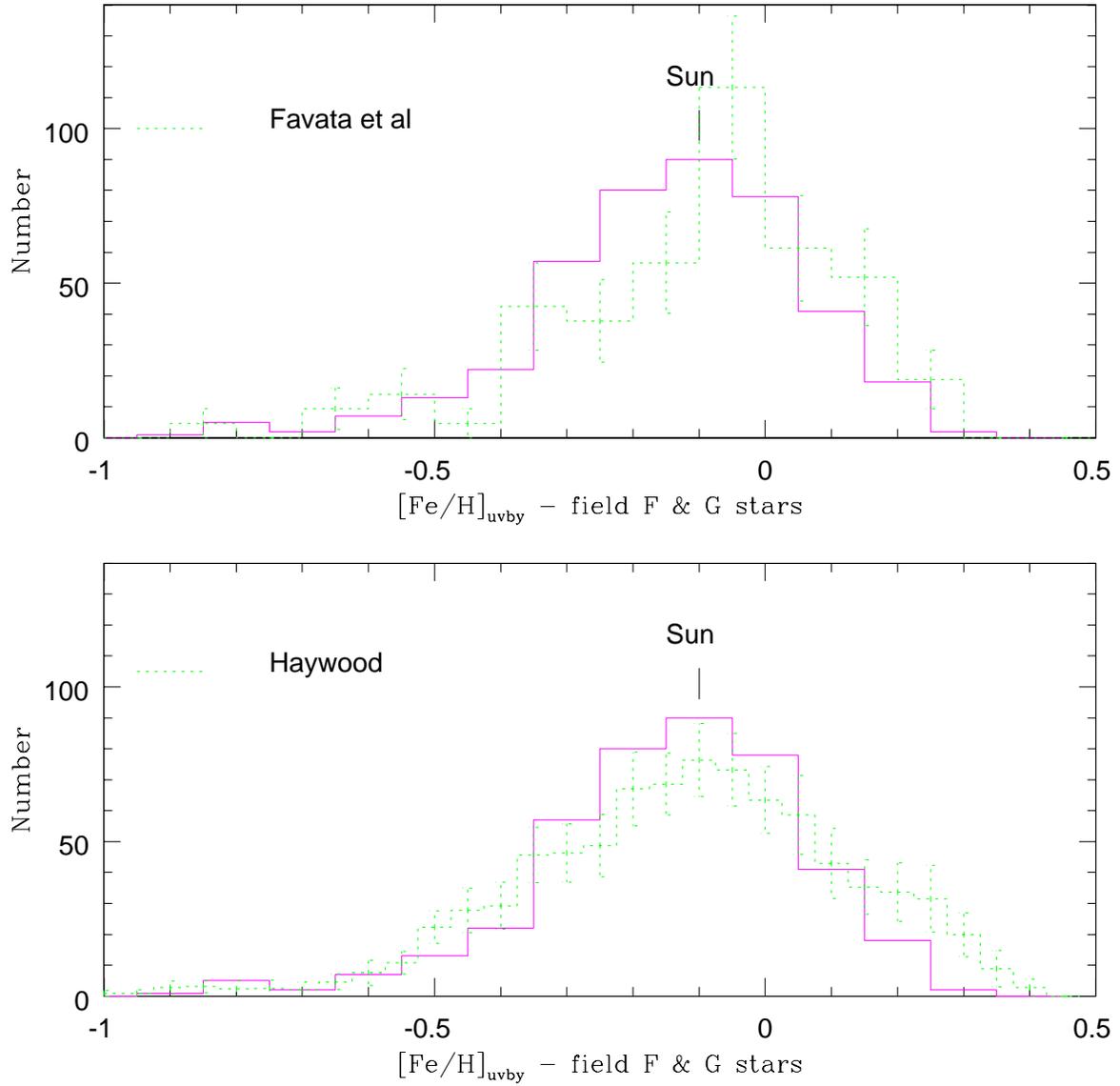}
\caption {Comparison of the abundance distributions derived for the FGK25 dataset and
the analyses by Favata {\sl et al.} (1997) and by Haywood (2001). See text for discussion.}
\label{fehhist}
\end{figure}

\begin{figure}
\figurenum{6}
\plotone{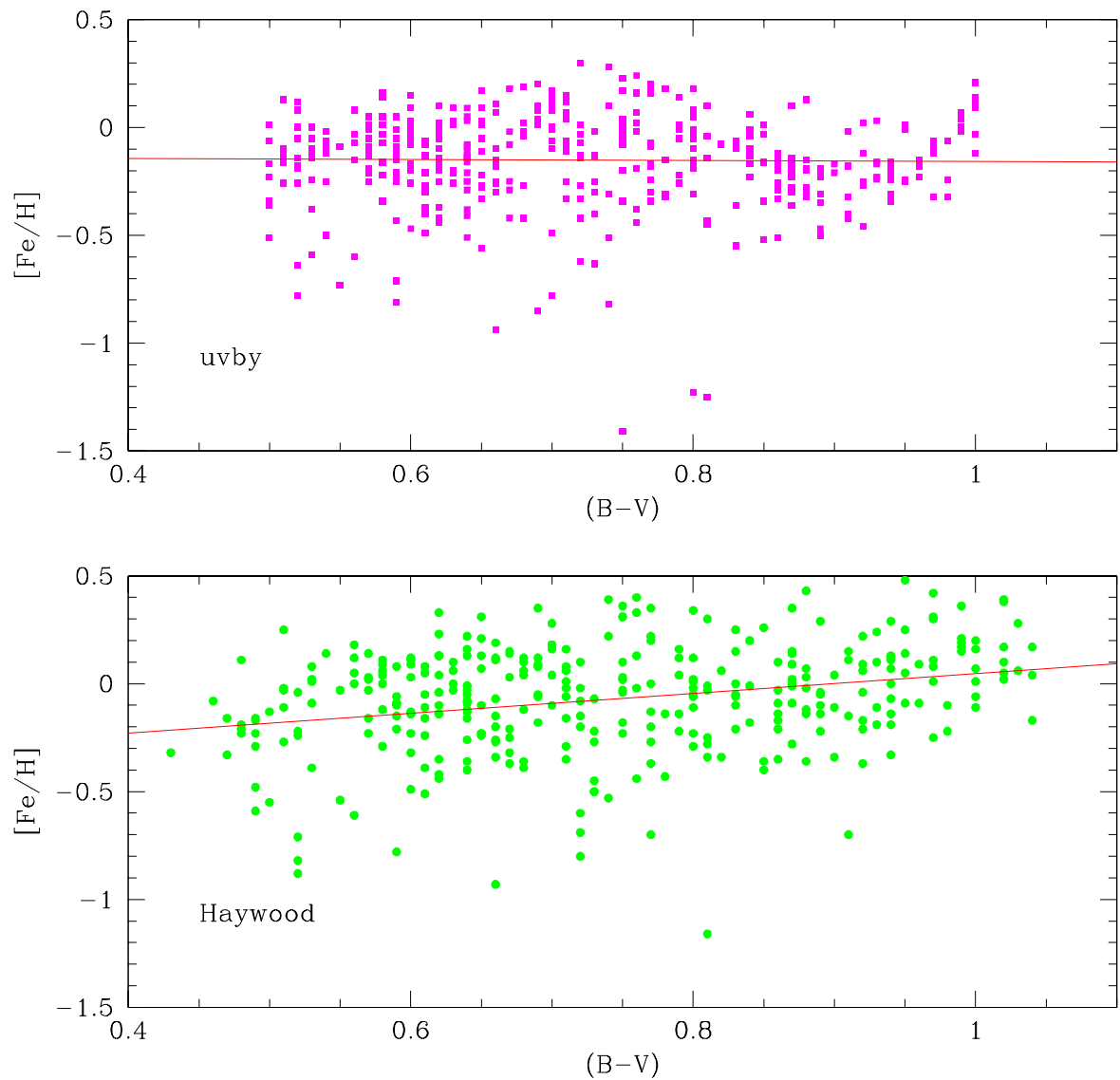}
\caption{ The abundance distribution as a function of (B-V) colour for stars in the FGK25
dataset and for the long-lived, M$_V > 4.5$ main-sequence stars in Haywood's (2001) analysis.
The latter stars show a clear trend, with $\langle [Fe/H] \rangle$ increasing with (B-V), 
suggesting a possible systematic error in the metallicity calibration of Geneva photometry.}
\label{bvfe}
\end{figure}

\begin{figure}
\figurenum{7}
\plotone{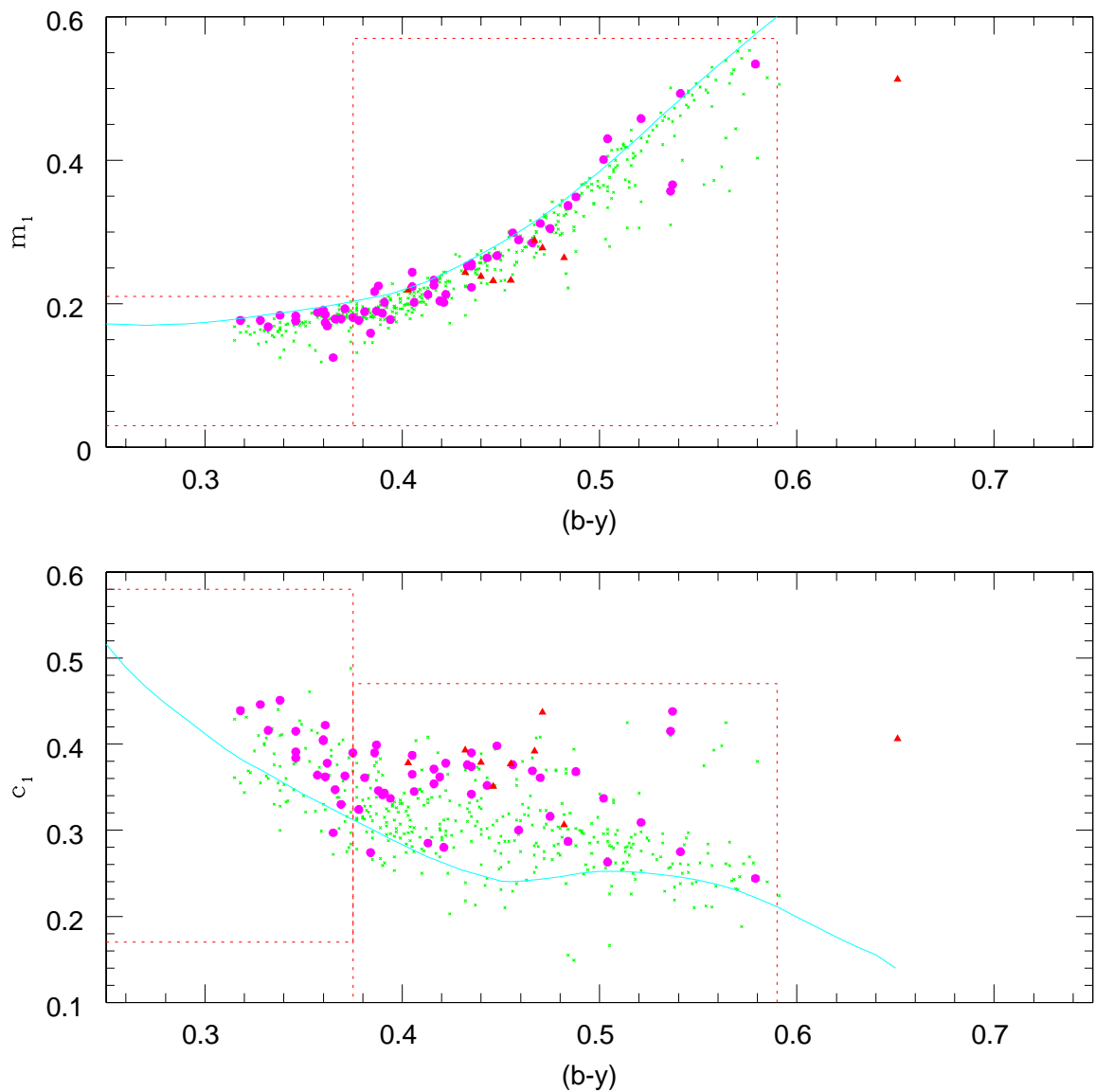}
\caption{ Str\"omgren data for the FGK25 Hipparcos sample (crosses) and the
ESP host stars (solid points). Stars from the latter sample identified as possible subgiants, 
based on their location in Figure 1, are plotted as solid triangles.
The dotted lines mark out the F-star and
G-star calibration r\'egimes from Schuster \& Nissen (1989). HIP 19221, or
HD 27442, is the only ESP host which lies outwith these limits. The solid line plots
the fiducial main-sequence, combining data from Davis Philip \& Egret (1980) and Olsen (1984). }
\label{uvby25g}
\end{figure}

\begin{figure}
\figurenum{8}
\plotone{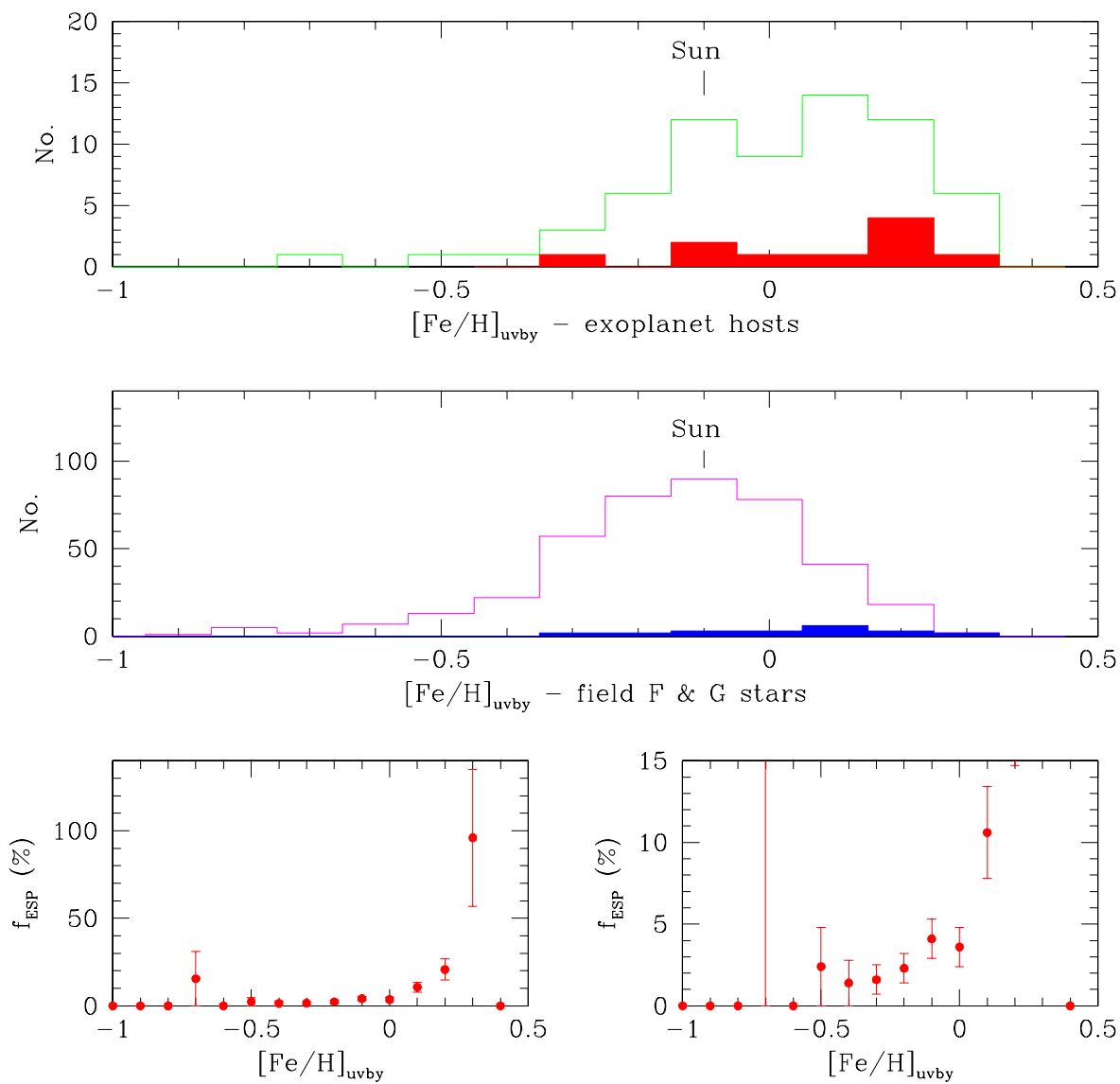}
\caption{ The upper panel plots the abundance distribution of the
ESP host stars, with solid histogram making the contribution from
stars identified as subgiants in Table 1. The middle panel plots the
comparable distribution for the  FGK25 sample. In both
cases, we use the Str\"omgren abundance calibration, which places the
solar abundance at approximately -0.1 dex. The lowest two panels show
the fraction of ESP host stars as a function of abundance, scaling the
full sample to 5\% of the FGK25 sample; the two panels plot identical
data, with the vertical scale extended in the  right hand panel.}
\label{fehcal}
\end{figure}

\begin{figure}
\figurenum{9}
\plotone{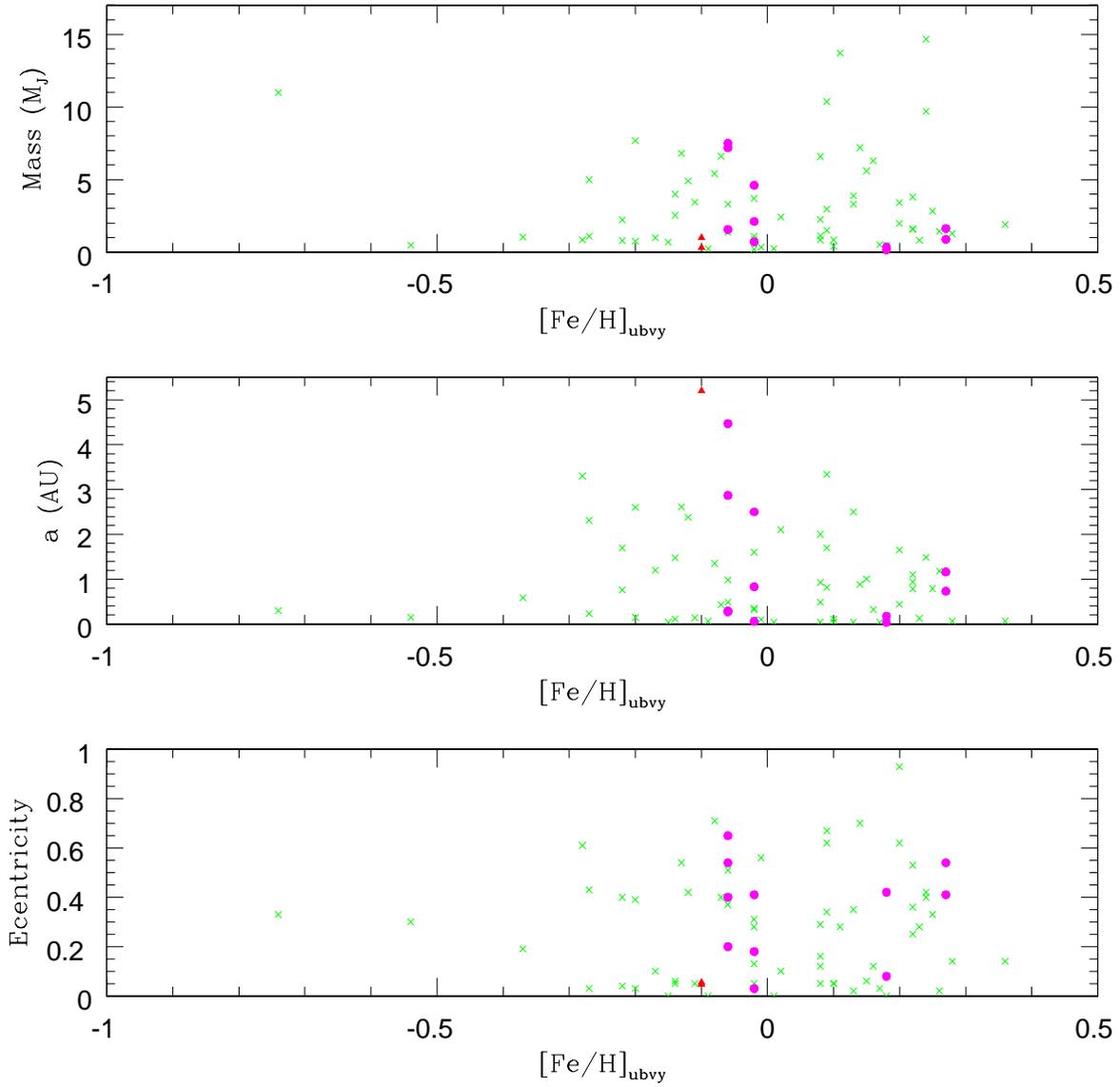}
\caption{ The distribution of properties of the known extrasolar planetary systems
as a function of abundance. Solid points identify systems with multiple planets; 
Jupiter and Saturn, representing the solar system, are plotted as solid triangles.
HD 114762b, the probable brown dwarf, is the most metal-poor object plotted here.}
\label{planref}
\end{figure}

\begin{figure}
\figurenum{10}
\plotone{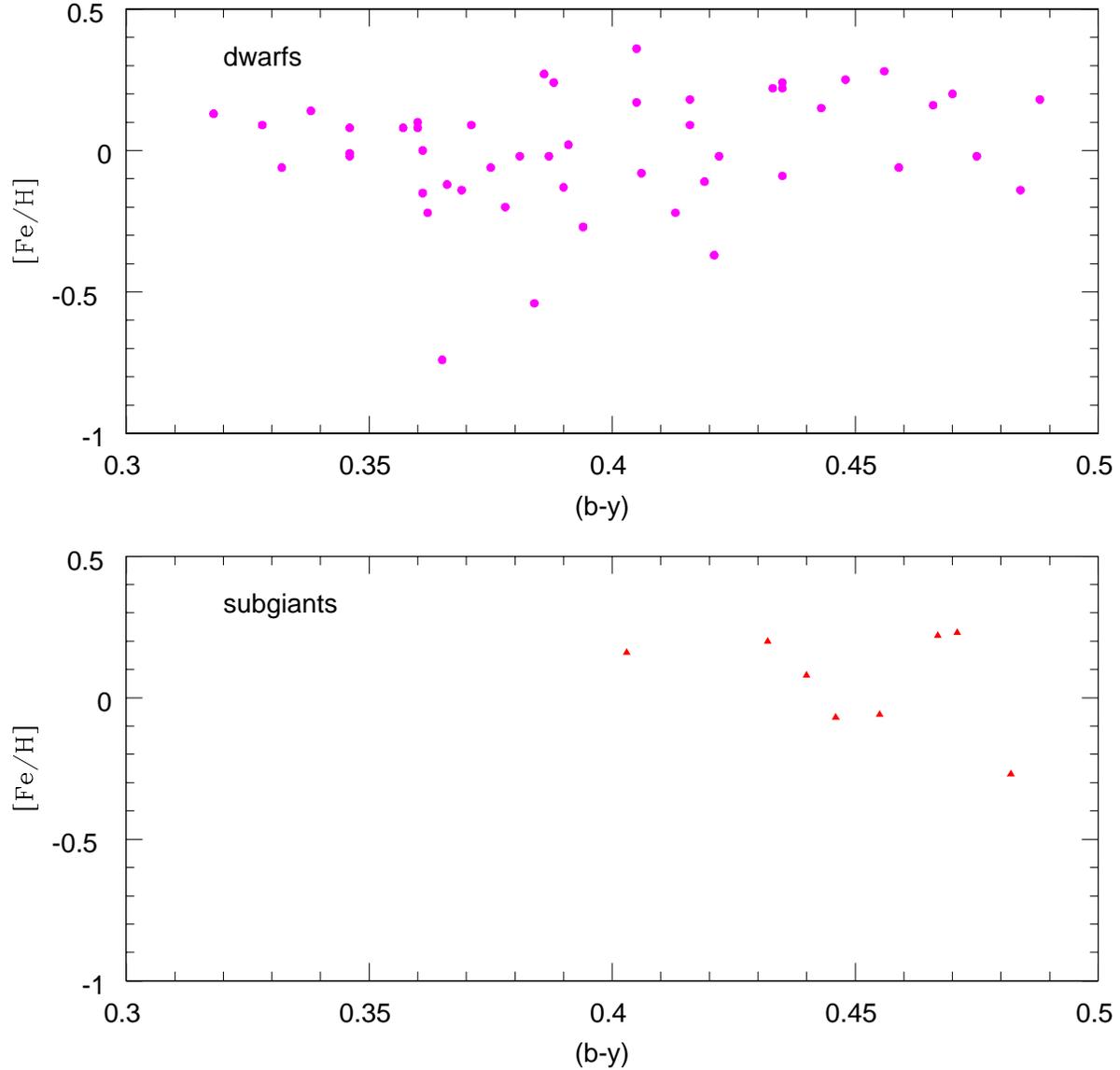}
\caption{ The metallicity distribution of the ESP hosts plotted as a function of 
(b-y) colour, separating stars classed as main-sequence and subgiants based on
their location in Figure 1. As Santos {\sl et al.} (2001) point out, there is
no evidence for a decrease in the maximum abundance as a function of colour
amongst the dwarfs; nor is there evidence for lower abundances amongst the
subgiants. Both results argue against the ESP host stars acquiring high metallicites
through planetary pollution.}
\label{fehdg}
\end{figure}

\begin{figure}
\figurenum{11}
\plotone{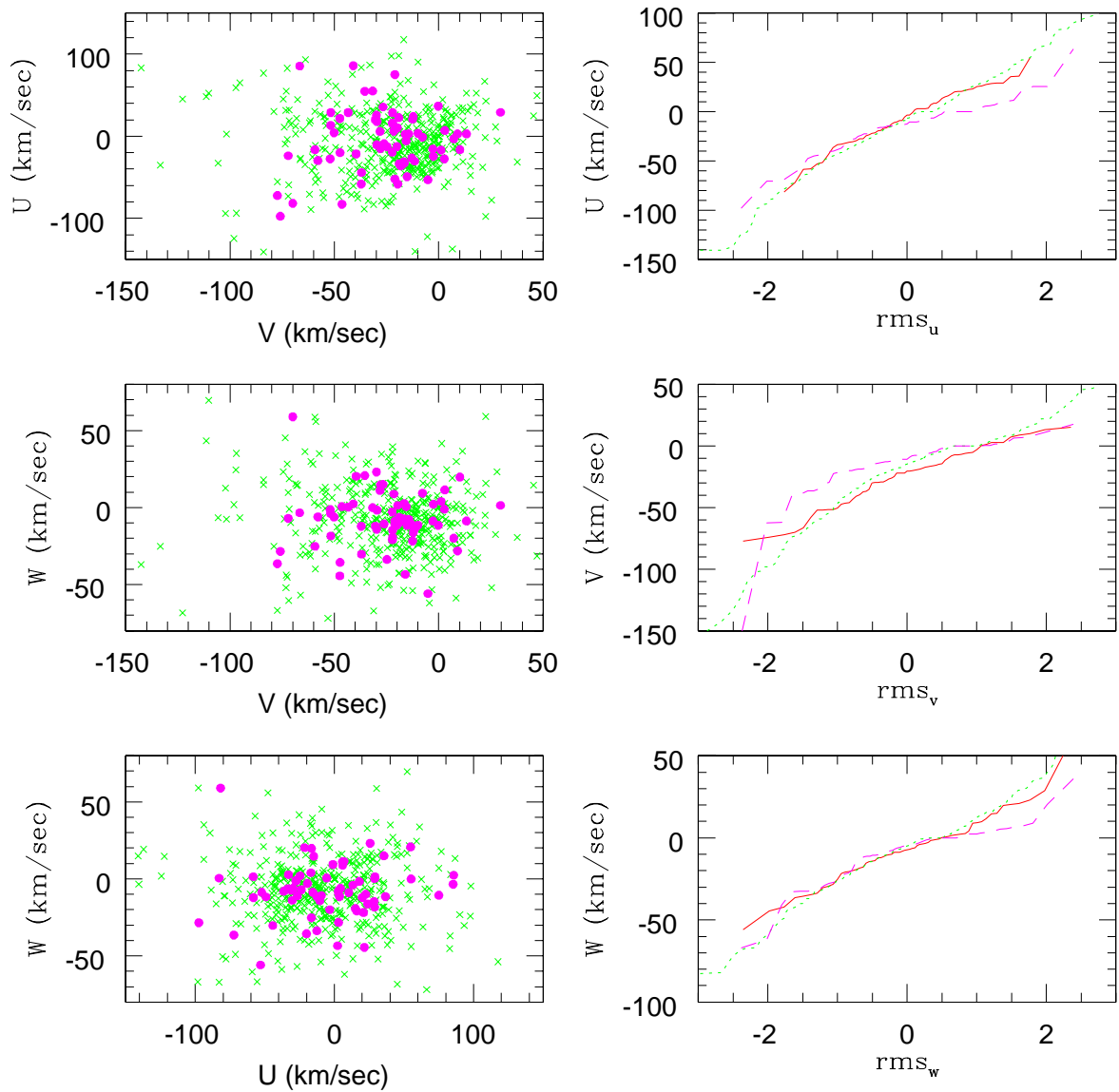}
\caption{Comparison between the space motions of the ESP hosts stars and local
disk dwarfs. The left-hand panels plot the two-component velocity distributions
for the ESP hosts (solid points) and the volume-complete M-dwarf sample (crosses) from the
PMSU. The right-hand panels show probability plots for the ESP hosts (solid line), PMSU
M dwarfs (dotted line) and dMe dwarfs (dashed line). A Gaussian distribution gives a
straight line, slope $\sigma$, in these diagrams.}
\label{uvwal}
\end{figure}

\begin{figure}
\figurenum{12}
\plotone{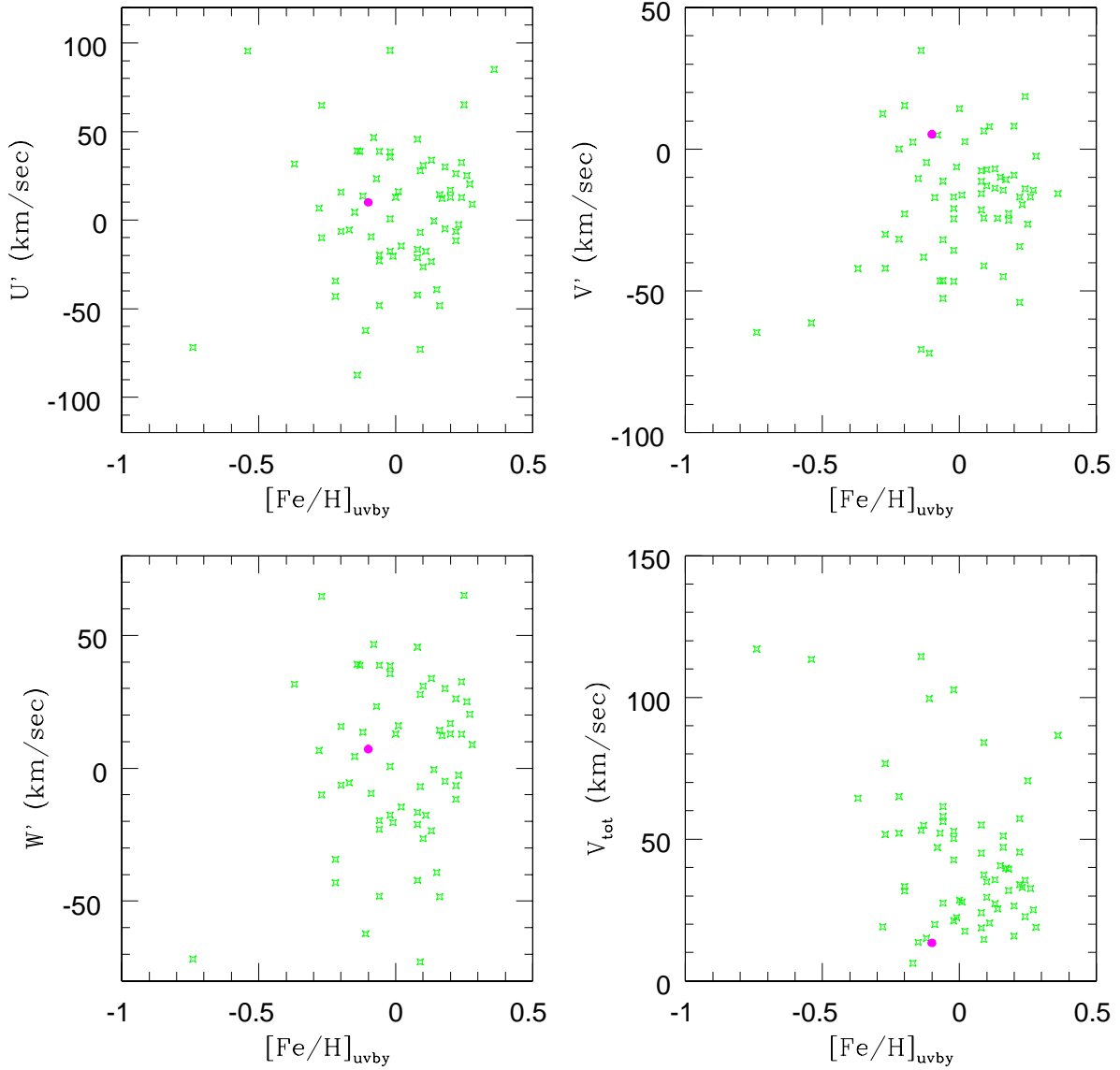}
\caption{The distribution of velocities of ESP host stars with respect to the Local Standard of Rest. The
Sun is plotted as a solid point.}
\label{uvfehpl}
\end{figure}

\begin{figure}
\figurenum{13}
\plotone{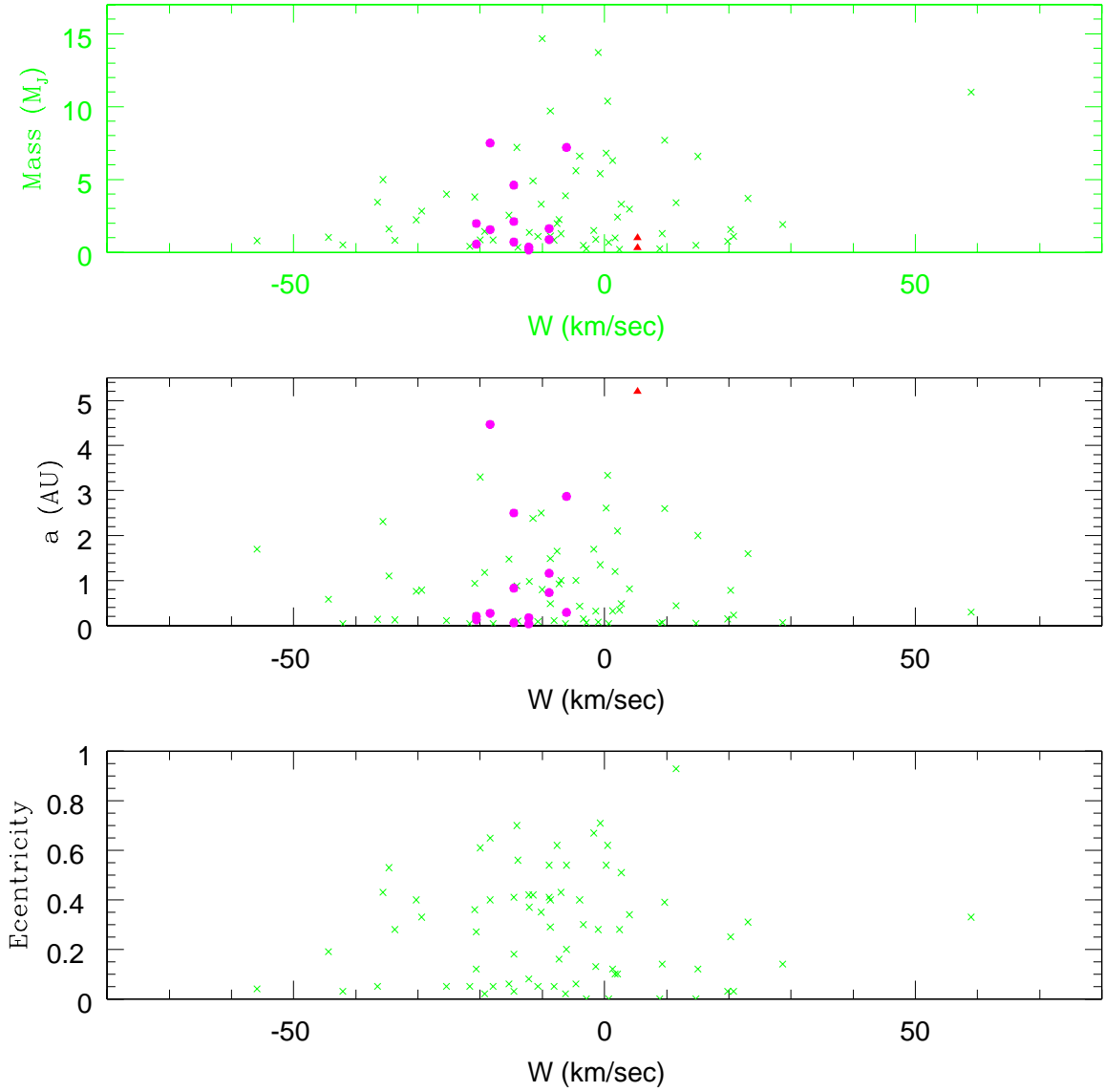}
\caption{The properties of the extrasolar planetary systems plotted as a function of W velocity.
The uppermost diagram,  plotting planetary mass (M$_2 \sin{i}$) against W, is the only comparison
which shows a suggestion of a trend, with higher-mass systems at low W velocities. } 
\label{planrefw}
\end{figure}


\begin{thebibliography}{}

\bibitem [Barrow \& Tipler, 1988] {bt88} Barrow, J.D., Tipler, F.J. 1988, {\sl The Anthropic Cosmological Principle}, Oxford Univ. press (Oxford)

\bibitem [Bessell, 2000] {b20} Bessell, M.S. 2000, PASP, 112, 961

\bibitem[Biemont et al, 1991] {biem} Biemont, E., Baudoux, M., Kurucz, R.L, Ansbacher, W.,
Pinnington, E.H. 1991, \aap, 249, 539

\bibitem[Butler et al, 2000] {bu20} Butler, R.P., Vogt, S.S., Marcy, G.W., Fischer, D.A.,
Henry, G.W., Apps, K. 2000, \apj, 545, 504

\bibitem [Cayrel de Strobel et al, 1997] {cds97} Cayrel de Stobel, G., Soubiran, C., Friel, E.D., Ralite, N., 
Francois, P. 1997, \aaps, 124, 299

\bibitem [Davis Philip \& Egret, 1980] {de80} Davis Philip, A.G., Egret, D. 1980, \aaps, 40, 199

\bibitem [Dehnen \& Binney, 1998] {db98} Dehnen, W., Binney, J.J. 1998, \mnras, 298, 387

\bibitem [Dick, 1998] {d98} Dick, S.J., 1998, {\sl Life on other worlds}, Cambridge University Press, 
(Cambridge)

\bibitem [Donahue, 1993] {d93} Donahue, R.A., 1993, Ph.d. thesis, New Mexico State University

\bibitem[Duflot et al, 1995]{du95} Duflot, M., Figon, P. Meyssonier, N. 1995, \aaps, 114, 269

\bibitem[ESA, 1997]{E97} ESA, 1997, The Hipparcos Catalogue

\bibitem[Fabricius \& Makarov, 2000] {fm20} Fabricius, C., Makarov, V.V. 2000, \aaps, 144, 45

\bibitem[Favata et al, 1996]{fav96} Favata, F., Micela, G., Sciortino, S. 1996, \aap, 311, 951

\bibitem[Favata et al, 1997]{fav97} Favata, F., Micela, G., Sciortino, S. 1997, \aap, 323, 809 (F97)

\bibitem[Fouts \& Sandage, 1986] {fs86} Fouts, G., Sandage, A. 1986, \aj, 91, 1189

\bibitem [Gatewood {\sl et al.}, 2001] {ha20} Gatewood, G., Han, I., Black, D.C. 2001, \apjl, 548, L61

\bibitem[Gimenez, 2000] {g20} Gim\'enez, A. 2000, \aap, 356, 213

\bibitem[Gizis {\sl et al.}, 2001] {giz20} Gizis, J.E., Reid, I.N., Hawley, S.L. 2001, \aj, 
subm. (PMSU3)

\bibitem[Gliese, 1969] {gl69} Gliese, W. 1969, Catalogue of Nearby Stars, 
Veroff. Astr. Rechen-Instituts, Heidelberg, Nr. 22 

\bibitem[Gliese \& Jahrei{\ss}, 1979] {gj69} Gliese, W.,  Jahrei{\ss}, H. 1979, A\&AS 38, 423

\bibitem [Gonzalez, 1997] {gon97} Gonzalez, G. 1997, \mnras, 285, 403

\bibitem[Gonzales, 1998] {gon98} Gonzalez, G. 1998, \aap, 334, 221

\bibitem [Gonzalez, 1999] {gon99} Gonzalez, G. 1999, \mnras, 308, 447

\bibitem[Gonzales \& Vanture, 1998] {gonv98} Gonzalez, G., Vanture, A.D. 1998, \aap, 339, L29

\bibitem[Gonzales et al, 1999] {gon1999} Gonzalez, G., Wallerstein, G., Saar, S. 1999,
\apjl, 511, L111

\bibitem[Gonzales \& Laws, 2000] {gc20} Gonzalez, G., Laws, C. 2000, \aj, 119, 390

\bibitem[Gonzales et al, 2001] {gon2001} Gonzalez, G., Laws, C., Tyagi, S., Reddy, B.E.
 2001, \aj, 121, 432

\bibitem [Gratton {\sl et al.}, 1997] {g97} Gratton, R.G., Carretta, E., Clementini, G., Sneden, C., 1997,
in {\sl Hipparcos Venice '97}, ed B Battrick (ESA), p 339

\bibitem[Griffin, 1972] {g72} Griffin, R.F. 1972, \mnras, 155, 449

\bibitem [Guillot, 1999] {gu99} Guillot, T. 1999, Science, 286, 272

\bibitem [Halbwachs {\sl et al.}, 2000] {hal20} Halbwachs, J.L., Arenou, F., Mayor, M., Udry, S., Queloz, D.
2000, \aap, 355, 581

\bibitem [Han {\sl et al.}, 2001] {ha20} Han, I., Black, D.C., Gatewood, G. 2001, \apjl, 548, L57

\bibitem[Hauck \& Mermilliod, 1998] {hm98} Hauck, B., Mermilliod, M. 1998, \aaps, 129, 431

\bibitem[Hawley et al, 1996] {haw96} Hawley, S.L., Gizis, J.E., Reid, I.N. 1996, \aj, 112, 2799 [PMSU2]

\bibitem[Haywood, 2001] {hay20} Haywood, M. 2001, \mnras, 325, 1365

\bibitem [Heacox, 1999]  {hea99} Heacox, W.D. 1999, \apj, 526, 928

\bibitem [Henry et al, 1998] {he98} Henry, T.J., Soderblom, D.R., Donahue, R.A., Baliunas, S.L. 1996, \aj, 111, 439

\bibitem [Jahrei{\ss} \& Wielen, 1997] {jw97} Jahrei{\ss}, H., Wielen, R. 1997, Proc. ESA Stmp. 402, {\sl Hipparcos - Venice '97},
ESA Publications, Noordwijk, p. 675

\bibitem [Jorissen, 2001] {j20} Jorissen, A, Mayor, M., Udry, S. 2001, \aap, 379, 992

\bibitem [Latham {\sl et al.}, 1989] {la89} Latham, D.W., Mazeh, T., Stefanik, R.P., Mayor, M., 
Burki, G. 1989, Nature, 339. 38

\bibitem[Laughlin, 2000] {l20} Laughlin, G. 2000, \apj, 545, 1064

\bibitem[Laws \& Gonzalex, 2001] {lg2001} Laws, C., Gonzalez, G. 2001, \apj, in press

\bibitem[Lin, Bodenheimer \& Richardson, 1996] {lbr96} Lin, D.N.C., Bodenheimer, P., Richardson, D.C.
1996, Nature, 380, 606

\bibitem[Lutz \& Upgren, 1980] {lu80} Lutz, T.E., Upgren A.R. 1980, \aj, 85, 573

\bibitem [McGrath {\sl et al.}, 2001] {mcg20} Mcgrath, M.A. {\sl et al.} 2001, DPS, 33, 6001

\bibitem [Marcy \& Benitz, 1989] {mb89} Marcy, G.W., Benitz, K.J. 1989, \apj, 344, 441

\bibitem[Marcy \& Butler, 2000] {mb2000} Marcy, G.W., Butler, R.P. 2000, \pasp, 112, 137

\bibitem[Mayor \& Queloz, 1995] {mq95} Mayor, M., Queloz, D. 1995, Nature, 378, 355

\bibitem [Murray {\sl et al.}, 2001] {m20} Murray, N., Chaboyer, B., Arras, P., Hansen, B., Noyes, R.W.
2001, \apj, 555, 810

\bibitem [Murray \& Chaboyer, 2001] {mb20} Murray, N., Chaboyer, B. 2001, \apj, subm.

\bibitem[Naef et al, 2001]{naef} Naef, D., Latham, D., Mayor, M. {\sl et al.}, 2001, \aap, 
in press

\bibitem[Nidever {\sl et al.}, 2002] {nid} Nidever, D., Marcy, G.W., Butler, R.P., Fischer, D.A., Vogt, S.S.
2002, \aj, subm.

 \bibitem[Olsen, 1984] {o84} Olsen, E.H. 1984, \aaps, 57, 443

\bibitem[Oppenheimer {\sl et al.}, 2001] {o20}  Oppenheimer, B.R., Hambly, N.C., Digby, A.P., Hodgkin, S.T.,
Saumon, D. 2001, Science, 292, 698

\bibitem [Pourbaix, 2001] {p20} Pourbaix, D. 2001, \aap, 369, L22

\bibitem [Pourbaix \& Arenou, 2001] {pa20} Pourbaix, D., Arenou, F. 2001, \aap, 372, 935

\bibitem[Randich et al, 1999] {r99} Randich, S., Gratton, R., Pallavicini, R., Pasquini, L., 
Carretta, E. 1999, \aap, 348, 487

\bibitem[Reid et al, 1995] {rhg95b} Reid, I.N., Hawley, S.L., Gizis, J.E. 1995, \aj, 110, 1838 [PMSU1]

\bibitem [Reid, 1998] {r98} Reid, I.N. 1998 \aj, 115, 204

\bibitem[Reid et al, 2002] {rhg20} Reid, I.N., Hawley, S.L., Gizis, J.E. 2002, \aj, subm. [PMSU4]

\bibitem[Reid et al, 2001] {r20} Reid, I.N., Sahu, K.C., Hawley, S.L. 2001, \apj, 559, 942

\bibitem[Ryan, 1992] {ry92} Ryan, S.G. 1992, \aj, 104, 1144

\bibitem[Santos et al, 2000] {san20} Santos, N.C., Israelian, G., Mayor, M. 2000, \aap, 363, 228

\bibitem[Santos et al, 2001] {san2001} Santos, N.C., Israelian, G., Mayor, M. 2001, \aap, 373, 1019

\bibitem[Sargent \& Beckwith, 1993] {sb93} Sargent, A.I., Beckwith, S.V.W. 1993, Physics Today, 
46, 22

\bibitem [Schaller et al] {sch92} Schaller, G.,  Schaerer, D., Meynet, G., Maeder, A. 1992, \aaps, 96, 269

\bibitem [Schaerer et al] {sch93} Schaerer, D., Charbonnel, C., Meynet, G., Maeder, A., Schaller, G.
1993, \aaps, 102, 339

\bibitem [Schuster \& Nissen, 1989] {sc89} Schuster, W.J., Nissen, P.F. 1989, A\&A, 221, 65

\bibitem [Soderblom {\sl et al.}, 1991] {s91} Soderblom, D.R., Duncan, D.K., Johnson, D.R.H. 1991, \apj, 375, 722

\bibitem [Spitzer \& Schwarzschild, 1953] {ss53} Spitzer, L., Schwarzschild, M. 1953, \apj, 118, 106

\bibitem [Stepinski \& Black] {stb} Stepinksi, T.F., Black, D.C. 2000, \aap, 356, 903

\bibitem [Str\"omgren, 1966] {str66} Str\"omgren, B. 1966,  Ann. Rev. Astr. Ap., 4, 433

\bibitem [Tabachnik \& Tremaine, 2001] {tt20} Tabachnik, S., Tremaine, S. 2001, \aj, in press

\bibitem[Tinney et al, 2001] {tin20} Tinney, C.G., Butler, R.P., Marcy, G.W., Jones, H.R.A., 
Penny, A.J., Vogt, S.S., Apps, K., Henry, G.W. 2001. \apj, 551, 507

\bibitem[Vogt et al, 2001a] {v2001a} Vogt, S.S., Butler, R.P, Marcy, G.W., Fischer, D.A., 
Pourbaix, D., Apps, K., Laughlin, G. 2001a, \apj, subm.

\bibitem[Vogt et al, 2001b] {v2001b} Vogt, S.S., Butler, R.P, Marcy, G.W., Apps, K. 2001b,
\apj, in press

\bibitem [Wielen, 1977] {w77} Wielen, R. 1977, \aap, 60, 263

\bibitem [Zucker \& Mazeh, 2001] {zm20} Zucker, S., Mazeh, T. 2001, \apj, 562, 1038

\bibitem[Zucker et al, 2001] {z2001} S. Zucker, D. Naef, D.W. Latham, M. Mayor, T. Mazeh,
{\sl et al.} 2001, \apj, in press



\end{thebibliography}
\end{document}